\documentclass{aa}  

\usepackage{graphicx}
\usepackage[varg]{txfonts}
\newcommand*{\Euclid}{\textit{Euclid}\xspace}

\newcommand*{\degree}{\si{\degree}}
\newcommand*{\arcminute}{\si{\arcminute}}
\newcommand*{\arcsecond}{\si{\arcsecond}}

\newcommand*{\hMpc}{\si{\hMpc}}

\newcommand*{\kmsMpc}{\si{\kmsMpc}}

\usepackage{csquotes}
\usepackage{natbib,twoopt}
\usepackage[breaklinks=true,pdfencoding=auto,psdextra]{hyperref} 
\bibpunct{(}{)}{;}{a}{}{,}             
\makeatletter
  \newcommandtwoopt{\citeads}[3][][]{\href{http://adsabs.harvard.edu/abs/#3}%
    {\def\hyper@linkstart##1##2{}%
     \let\hyper@linkend\@empty\citealp[#1][#2]{#3}}}
  \newcommandtwoopt{\citepads}[3][][]{\href{http://adsabs.harvard.edu/abs/#3}%
    {\def\hyper@linkstart##1##2{}%
     \let\hyper@linkend\@empty\citep[#1][#2]{#3}}}
  \newcommandtwoopt{\citetads}[3][][]{\href{http://adsabs.harvard.edu/abs/#3}%
    {\def\hyper@linkstart##1##2{}%
     \let\hyper@linkend\@empty\citet[#1][#2]{#3}}}
  \newcommandtwoopt{\citeyearads}[3][][]%
    {\href{http://adsabs.harvard.edu/abs/#3}
    {\def\hyper@linkstart##1##2{}%
     \let\hyper@linkend\@empty\citeyear[#1][#2]{#3}}}
\makeatother

\usepackage{breakurl} 

\usepackage{newtxtext,newtxmath}

\usepackage[T1]{fontenc}
\usepackage{ae,aecompl}

\usepackage{graphicx}	
\usepackage{amsmath}	
\usepackage{xspace}
\usepackage{amsfonts,textcomp}
\usepackage[usenames]{color}
\usepackage{times}

\hypersetup{
    colorlinks=true,
    linkcolor=blue,
    filecolor=magenta,      
    urlcolor=blue,
    citecolor=blue
}
\usepackage{subfigure}
\usepackage{ulem}
\usepackage{enumitem}
\usepackage[compatibility=false]{caption}
\usepackage{subcaption}
\usepackage{multirow}
\usepackage{tabulary}
\usepackage[para]{threeparttable}
\usepackage{array,booktabs,longtable,tabularx}
\newcolumntype{L}{>{\raggedright\arraybackslash}X}
\usepackage{ltablex}
\usepackage{comment}

\graphicspath{{../},{./}}
\usepackage{amsmath}

\definecolor{Orange}{rgb}{1.0,0.5,0.15}
\definecolor{Blue}{rgb}{0,0.08,0.65}
\definecolor{Blue2}{rgb}{0,0.4,0.6}

\definecolor{Red}{rgb}{0.65,0.08,0.05}
\definecolor{Green}{rgb}{0.15,0.45,0.25}
\definecolor{Pink}{rgb}{1.0,0.05,0.5}
\definecolor{Purple}{rgb}{0.3,0.,0.5}

\usepackage[switch, modulo]{lineno}
\begin{document}


   \title{Reconstructing galaxy star formation histories from COSMOS2020 photometry  using simulation-based inference}

   
    \author{G.~Aufort\inst{1}, C.~Laigle\inst{1}, H.~J.~McCracken\inst{1}, D.~Le~Borgne\inst{1}, R.~Arango-Toro\inst{2}, L.~Ciesla\inst{2}, O.~Ilbert\inst{2}, L.~Tresse\inst{2} and
    Y.~Dubois\inst{1}
    }

    \institute{	
    \inst{1}CNRS and Sorbonne Universit\'e, UMR 7095, Institut d'Astrophysique de Paris, 98 bis, Boulevard Arago, F-75014 Paris, France\\
    \inst{2}Aix Marseille Univ., CNRS, CNES, LAM, Marseille, France
    }	

\date{Accepted XXX. Received YYY; in original form ZZZ}

 \abstract
{
We propose a novel method to reconstruct the full posterior distribution of the star formation histories (SFHs) of galaxies from broad-band photometry. Our method combines the simulation-based inference (SBI) framework using a neural network trained with SFHs and photometry from the {\sc Horizon-AGN} hydrodynamical cosmological simulation. We apply our technique to reconstruct SFHs in the COSMOS Treasury field using only COSMOS2020 photometry in the redshift range $0<z<3$. The method is able to accurately estimate the SFH and quantify the Bayesian uncertainty on simulated data, with an unbiased posterior mean, $\sigma_{\rm err}\leq 0.16$ dex for all formation times and properly calibrated posterior intervals.  Our SFHs are in broad agreement with literature measurements derived by different methods using combined photometric and spectroscopic datasets. 
The SFHs of galaxies as a function of location in the $\mathrm{NUV}-r$ versus $r-J$ colour-colour diagram are in general agreement with expectations, varying smoothly from star-forming to passive and quiescent galaxies being well localized in the red part of the diagram. We extract summary statistics to quantify the shape of the SFH, number of peaks, and formation redshift. The slopes of the SFHs of passive galaxies show only a weak trend with stellar mass at $z<1.35$ but a significant scatter, indicating that other factors than mass could drive the suppression of star-formation. Nevertheless, star-forming galaxies show a clear mass-dependent SFH, with lower-mass galaxies undergoing more vigorous recent star-formation. Overall, SFH slopes in COSMOS vary over a wider range than in {\sc Horizon-AGN}.
 Low-mass galaxies have more peaks in their mass assembly histories than high-mass ones, and the trend is clearer in COSMOS than in {\sc Horizon-AGN}. At a given mass, we find many different formation redshifts, but for passive galaxies the mass dependency of formation redshifts is weak. Most passive galaxies with stellar mass $\log M_*/M_\odot > 9$ had a first event of mass assembly around $z\sim 3$ ($2.2<z<5.8$), independent of mass.
This work represents a pilot study for the future analysis of the \textit{Euclid} Deep fields that  will reach similar depths in alike set of photometric bands, but with over an order-of-magnitude larger area, opening the possibility of deriving SFHs for millions of galaxies in a robust manner. }


   \authorrunning{Aufort et al.}
   \titlerunning{SFH reconstruction}
   
   \maketitle
   


\section{Introduction}
With the new avalanche of wide and deep astronomical surveys, like ESA's \textit{Euclid} survey \citep{euclid}, astronomers now face an increasingly difficult challenge to integrate the many pieces of the galaxy formation puzzle into a comprehensive model. 
 How can we build a consistent scenario of galaxy evolution? One of the difficulties is that it is impossible  to observe the evolution of a given galaxy: the best that we can do is to consider galaxy populations at different cosmic times and then try to connect them together with physical models. For example, the evolution of mass functions of passive and star-forming galaxy populations classified from a colour-colour diagram is a first-order approach to understand the rise-up and quenching of star formation in the Universe \citep[e.g.][amongst others]{Peng2010,2014ARA&A..52..415M,santini22,weaver23, Gould2023}. In this approach, only the mass and star-formation rate of the galaxies at the observation redshift is considered, mixing galaxies with potentially different star formation and quenching histories and making difficult the physical interpretation of the mass function evolution.
 
Galaxy spectra are the cumulative emission and absorption of the many stellar populations which constitute them. Those populations have various age and metallicity depending on their formation times and conditions and are susceptible to imprint specific signatures in galaxy spectra.
Therefore the build-up history of galaxies is encoded to some extent in their photometry. A technique to automatically extract individual star-formation histories from photometric data could be a revolutionary way to improve our understanding of galaxy formation and mass build-up in the Universe. This would allow to sort galaxies not only by their mass and star-formation rates, but also using specific metrics based on the star formation histories (SFH) (e.g. the formation redshift, the number of bursts, the slope of the SFHs in their recent history) which are more directly connected to signature of physical processes impacting galaxy histories such as feedback and environment. Unfortunately, systematically reconstructing such histories from broad-band photometry alone has always been challenging with commonly used template-fitting codes originally designed for photometric redshift estimation (e.g. {\texttt{LePHARE}} \citealp{arnouts02,ilbert06}, {\tt EAZY} \citealp[]{brammer08} amongst others).  In fact, even the reconstruction of the recent star-formation rate is questionable from standard spectral energy distribution (SED)-fitting \citep[e.g.][]{laigle19}. This is because computing SFHs from large datasets of intermediate to low signal-to-noise photometry is limited by the sparsity of the model set. Furthermore, large samples of photometric measurements often have only a few broad-band filters.  A key issue is therefore to optimally exploit the entirety of the information contained in broad-band colours, while fitting physically motivated SFH models.

To improve the results of these classical template-fitting codes, one could think about increasing the size of the template library, by generating SED from many different SFHs through more sophisticated parametrisation \citep[e.g.][]{pacifici16}. These techniques are however computationally expensive and cannot be blindly applied to billions of galaxies. In addition, choosing the correct models to use is challenging as it significantly impacts the inference and justifying the choice of one model over another is problematic.
Another exploration avenue to be mentioned involves simpler analytical forms but optimised to track only specific features in SFHs using high signal-to-noise ratio photometry. Although efficient, by design these methods suffer from a lack of generality since they are limited to the identification of specific populations, such that those which undergo a recent dramatic event \citep[e.g.][]{Ciesla18,Aufort2020}.

To circumvent the difficulties inherent to the use of parametric SFH models \citep[see also][]{Carnall19}, some have explored the use of  non-parametric SFHs that can be implemented in SED-fitting codes (e.g. {\tt Prospector} \citealp{prospectorref} or {\tt CIGALE} \citealp{boquien19}), as e.g. presented in \cite{Iyer19},  \cite{Tacchella22b}, \cite{ciesla23}, \cite{wang24}, or Arango et al. in prep. In general these methods outperform traditional parametric methods, but are still inevitably dependent on the choice of the prior (e.g. continuous or stochastic) for the non-parametric model \citep{leja19a,Suess22}.
In this context, machine learning approaches have also gained momentum due to their ability to optimally exploit all information contained in multidimensional datasets. Moreover, recent advances in statistical machine learning techniques \citep{NIPS2016_6aca9700} formalised the link between machine learning methods and Bayesian inference. 

In this paper, we propose a new technique to bypass the limitations of traditional SED fitting approaches by using a simulation-based inference (SBI) framework to reconstruct the SFHs of galaxies. Our technique relies on the hydrodynamical cosmological simulation {\sc Horizon-AGN} \citep{dubois14} as a realistic SFH model. Our model can be parametrised in a very flexible and effective way, from which Bayesian inference can be performed to estimate the SFH of a galaxy using only photometry. For our photometric data, we use the COSMOS2020 photometric catalogue \citep{weaver22}, which draws on the vast amount of photometric data present in the COSMOS field \citep{cosmos}. COSMOS also contains spectroscopic data which we can use as an independent validation of our method. 

Looking forward, the \Euclid mission \citep{euclid,EuclidSkyOverview} promises to revolutionise not only our knowledge of the cosmological model but also of galaxy formation and evolution. The \Euclid survey, now underway since February 2024, will comprise a $\sim15000\deg^2$ wide survey \citep{Scaramella22} together with deep survey with forty times the exposure of the wide survey. These \Euclid deep fields will contain, crucially, very deep near-infrared photometry from \textit{Euclid}'s Near-Infrared Spectrograph instrument \citep{euclid,Scaramella22,Euclid2024}, optical data from Hyper-Suprime-Cam on Subaru \citep{euclidcollaboration2024euclidpreparationcosmicdawn} and infrared data from \textit{Spitzer}'s IRAC camera \citep{moneti22}. These data represent the only survey, currently or in planning, that is capable of making stellar-mass-selected samples at these depths and areas. These data will be an ideal sample for investigating galaxy SFHs in different populations and probe the currently unexplored ones. In turn, we will be able to determine what triggers dramatic changes in galactic SFH, when each of these processes occurs, and where in the large-scale structure they are the most efficient at transforming galaxies. One of the primary objectives of this paper is to understand how well SFHs can be reconstructed with photometry similar to the \Euclid deep fields. {We note that the filter set used here is slightly different from the one that is obtained on the \Euclid deep fields, but our aim is not to determine the optimal filter set for the SFH reconstruction. We rather want to demonstrate that such reconstruction is possible from photometry with a limited number of filters.}

Section~\ref{sec:data} presents our photometric catalogue together with the cosmological simulation used as a model for the SFH inference. Section~\ref{sec:method} describes our SBI method. In Section~\ref{sec:results} we present our results and consistency checks with the simulations and with recent work from the literature. Section~\ref{sec:caveats} highlights caveats and perspectives. Our conclusions are presented in Section~\ref{sec:conclusion}.

We use the WMAP7 cosmology~\citep{komatsuetal11} for our inference pipeline (as is used in {\sc Horizon-AGN}). In COSMOS, galaxy properties were estimated by \texttt{LePHARE} assuming Planck13 cosmology \citep{Planck2013} but the effect of changing cosmology is not significant considering the other sources of uncertainties. We use AB magnitudes \citep{1983ApJ...266..713O} throughout. 
\section{Data and model}
\label{sec:data}

\subsection{Observations, photometry, photometric redshifts}

The COSMOS field has amongst the most extensive multiwavelength coverage of any extragalactic survey. In this work, we use a subset of the rich COSMOS photometry with the aim of approximating the filter set and depths which will be available in the \Euclid deep survey. Specifically, we restrict ourselves to $u^*$ from the CLAUDS survey \citep{sawicki19}, $g$, $r$ $i$ and $z$ from HSC, $Y$, $J$, $H$ and $K_{\rm s}$ from UltraVISTA \citep{mccracken12}. 

Our photometric measurements come from the \textsc{ Classic} version of the COSMOS2020 catalogue \citep{weaver22}. This catalogue uses \texttt{SExtractor} \citep{bertin96} to measure galaxy colours in 2\arcsec apertures. The photometry reaches a $3\sigma$ depth of 28.1 mag and 25.3 in 2$\arcsec$~apertures in $g$-band and $K_{\rm s}$-bands respectively. We limit our sample at ${\rm SNR}>2$ in the $K_{\rm s}$-band, and restrict ourselves to the $z<3$ redshift range. This ensures that the available photometric set is relatively homogeneous over the optical rest-frame. Finally, only objects within the unmasked area\footnote{using the COSMOS2020 {\tt FLAG\_COMBINED} column} are kept in the sample. In addition, by design, objects with photometry missing in at least one of the bands will be removed from the sample. The total area after conservative masking is 1.27~deg$^2$ \citep{weaver22}. The final area has coverage in HSC, and UltraVISTA and is not affected by artefacts or bright stars.

For the analysis of our reconstructed SFHs (but not for their reconstruction itself), we rely on photometric redshifts, stellar masses and absolute magnitudes derived from SED fitting with \texttt{LePHARE} over the COSMOS photometric bands \citep{arnouts02,ilbert06} using the same configuration as \cite{ilbert13}.  As an example of the performance of this code in our sample, the precision and fraction of outliers of photometric redshifts is better than 1\% in the brightest bin $i<22.5$, and is still of the order of $\sim$4\% in the faintest bin $25<i<27$, with a fraction of outliers of $\sim$20\%. 

\subsection{Galaxy classification}

An important aim of this paper is to investigate the dependence of the SFH on stellar mass and star-forming types by subdividing the parent population according to these properties (see Section~\ref{sec:results}). To do so, we rely on the physical parameters in the COSMOS2020 catalogue. First, masses are those derived from the \texttt{LePHARE} SED fitting. Systematic errors in galaxy masses derived this way are generally small, below $\sim 0.15$~dex and the scatter is generally smaller than $\sim 0.1$~dex \citep{laigle19}. We also classify galaxies as passive or star-forming based on their location in $\mathrm{NUV}-r$ versus $r-J$ colour-colour diagram \citep{ilbert13}. Given the photometric redshift of a galaxy, rest-frame magnitudes are estimated while relying on the apparent magnitude probing the part of the galaxy spectrum which is the closest to the rest-frame filter. This approach allows to minimise the impact of the $k$-correction \citep{ilbert05}. 

\subsection{The model: The {\sc Horizon-AGN} simulation}
\label{sec:hzagn}
The {\sc Horizon-AGN} simulation\footnote{\url{http://www.horizon-simulation.org/}} \citep{dubois14} is a cosmological hydrodynamical simulation run in a ($100\,h^{-1}\,\rm Mpc$)$^3$ box with the adaptive mesh refinement code {\texttt{ramses}}~\citep{teyssier02}. It adopts a $\Lambda$CDM cosmology with total matter density $\Omega_{\rm  m}=0.272$, dark energy density $\Omega_\Lambda=0.728$, matter power spectrum amplitude $\sigma_8=0.81$, baryon density $\Omega_{\rm  b}=0.045$, Hubble constant $H_0=70.4 \, \rm km\,s^{-1}\,Mpc^{-1}$, and $n_{\rm s}=0.967$ compatible with the WMAP-7 data~\citep{komatsuetal11}. The initially coarse $1024^3$ grid is adaptively refined based on density down to $1$ physical kpc. The volume contains $1024^3$ dark matter particles, corresponding to a dark matter mass resolution of $M_{\rm  DM, res}=8\times 10^7 \, M_\odot$. When it was generated, {\sc Horizon-AGN} was implementing our state-of-the-art knowledge of galaxy physics which still hold now, including gas cooling and heating, star formation and stellar feedback, black hole formation, growth and feedback from active galactic nuclei \citep{dubois14}. In similar mass and redshift ranges, measurements from {\sc Horizon-AGN} are in relatively good agreement with observed cosmic star-formation histories, and mass and luminosity functions. 

However, some difficult-to-resolve discrepancies remain. In particular, low-mass simulated galaxies form too many stars at early times. The resulting stellar mass functions \citep{kaviraj17,picouet2023} and the stellar-to-halo mass relation are overestimated at the low-mass end \citep{shuntov22}. In addition, the fraction of high-mass quenched galaxies is smaller in the simulation compared to observations because  in general there is always some residual star formation in massive galaxies \citep{dubois16}.
For morphologies, \textsc{Horizon-AGN} correctly reproduces the observed morphological diversity \citep{dubois16} with the limitation that low-mass galaxies (those below $10^{9}\,M_\odot$) are unresolved.

\subsubsection{Star formation histories of simulated galaxies}
\label{sec:sfhzagn}
We use a photometric catalogue built from the central 1~deg$^2$  of the light cone that has been built ``on-the-fly'' from the simulated box. Gas, stars, and dark matter particles are extracted at each coarse time step according to their proper distance to the observer at the origin. In total, the light cone contains about 22000 portions of concentric shells. Galaxies were identified by running {\sc AdaptaHOP} \citep{aubert04} on the stellar particle distribution and identifying structures with a density threshold equal to 178 times the average matter density at that redshift. The local stellar particle density was computed from the 20~nearest neighbours. 

For all simulated galaxies, the age distribution of stellar particles belonging to these galaxies is known. From this we reconstruct the SFHs by summing the current stellar mass formed per bin of look-back time. Obviously star formation in the simulated galaxies is discrete, due to the mass and time resolution of the simulation In this definition, the stellar particle masses are not corrected for the stellar mass loss that occurs due to supernova ejecta and stellar winds.  
\footnote{In {\sc Horizon-AGN}, most of the stellar mass losses occur on a relatively short timescale ($\sim 100$~Myr, see e.g. Figure~A5 in \citealp[]{laigle19}), so the difference between the simulated SFH corrected or not from the stellar mass losses will mostly concern the amount of mass in our last SFH bin, and could change it by $\sim 10\%$).}

In the SFHs, we cannot distinguish between the stellar mass formed in situ, in the main progenitor, and with the stellar brought ex situ by accretion of satellite galaxies. Therefore, the SFHs include all the mass gained due to merger histories, and, consequently, a peak in the SFH could correspond either to a burst of star formation in the main progenitor or not, and could be triggered by mergers or in isolation.
It would therefore be more appropriate to talk about the ``mass assembly history'' rather than the ``star formation history''.  
 
\begin{figure}
 \includegraphics[width=\columnwidth]{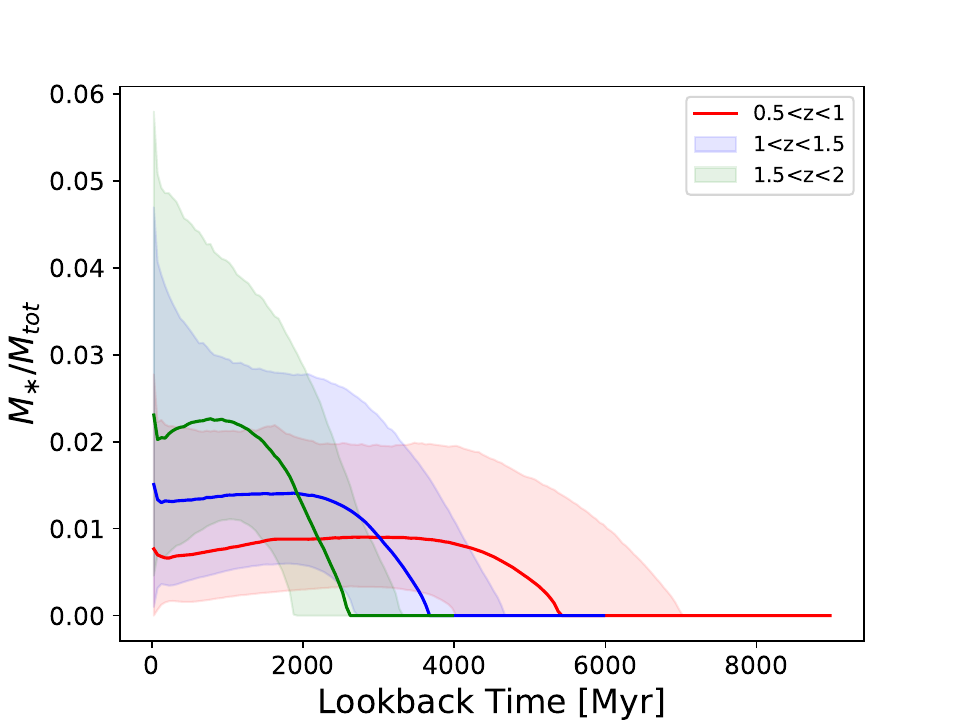}
 \caption{Average distributions of SFHs in {\sc Horizon-AGN}. Ribbons show $95\%$ intervals.}
 \label{fig:SFH_hz}
\end{figure}

 Fig.~\ref{fig:SFH_hz} presents the average shape of the SFH in {\sc Horizon-AGN} per bin of redshifts. SFHs are normalised by the total mass of galaxies at the time of observation (i.e. look-back time $=0$).

\subsubsection{Preparing the photometric training set from the simulation}
We computed galaxy photometry using the technique outlined in \cite{laigle19} and \cite{cadiou24}. This method assumes a \cite{chabrier03} initial mass function, single stellar population models from \cite{bruzual&charlot03}, and a $R_V = 3.1$ Milky Way dust grain model from \cite{weingartner01}. Dust attenuation is modelled along the line-of-sight of each star particle, using the gas-metal mass distribution as a proxy for dust distribution, and assume a fixed dust-to-metal mass ratio of 0.3.  The photometric flux errors are then added to the fluxes. {Specifically, the error is added in a given filter by randomly perturbing the flux. Galaxy flux error are a combination of a gaussian error with a standard deviation defined according to the COSMOS2020 depth in this filter and Poisson error (using the appropriate instrumental gain}). {Because of the simulation resolution limit, this simulated sample is limited to galaxies more massive than $10^9\,M_\odot$.}
\section{Methodology}
\label{sec:method}

\begin{figure*}
 \includegraphics[width=\textwidth]{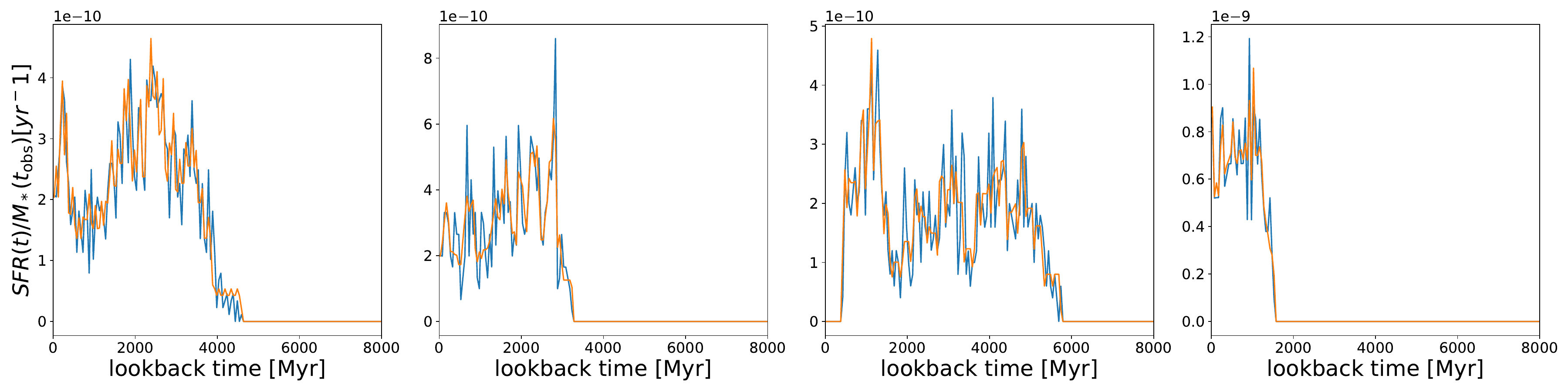}
 \caption{A few examples of the effect of our parametrisation on our {\sc Horizon-AGN} galaxies: in blue the original SFH binned linearly, and in orange the reconstruction after interpolating between 30 mass quantiles.}
 \label{fig:binning_hz}
\end{figure*}

\begin{figure}
 \includegraphics[width=\columnwidth]{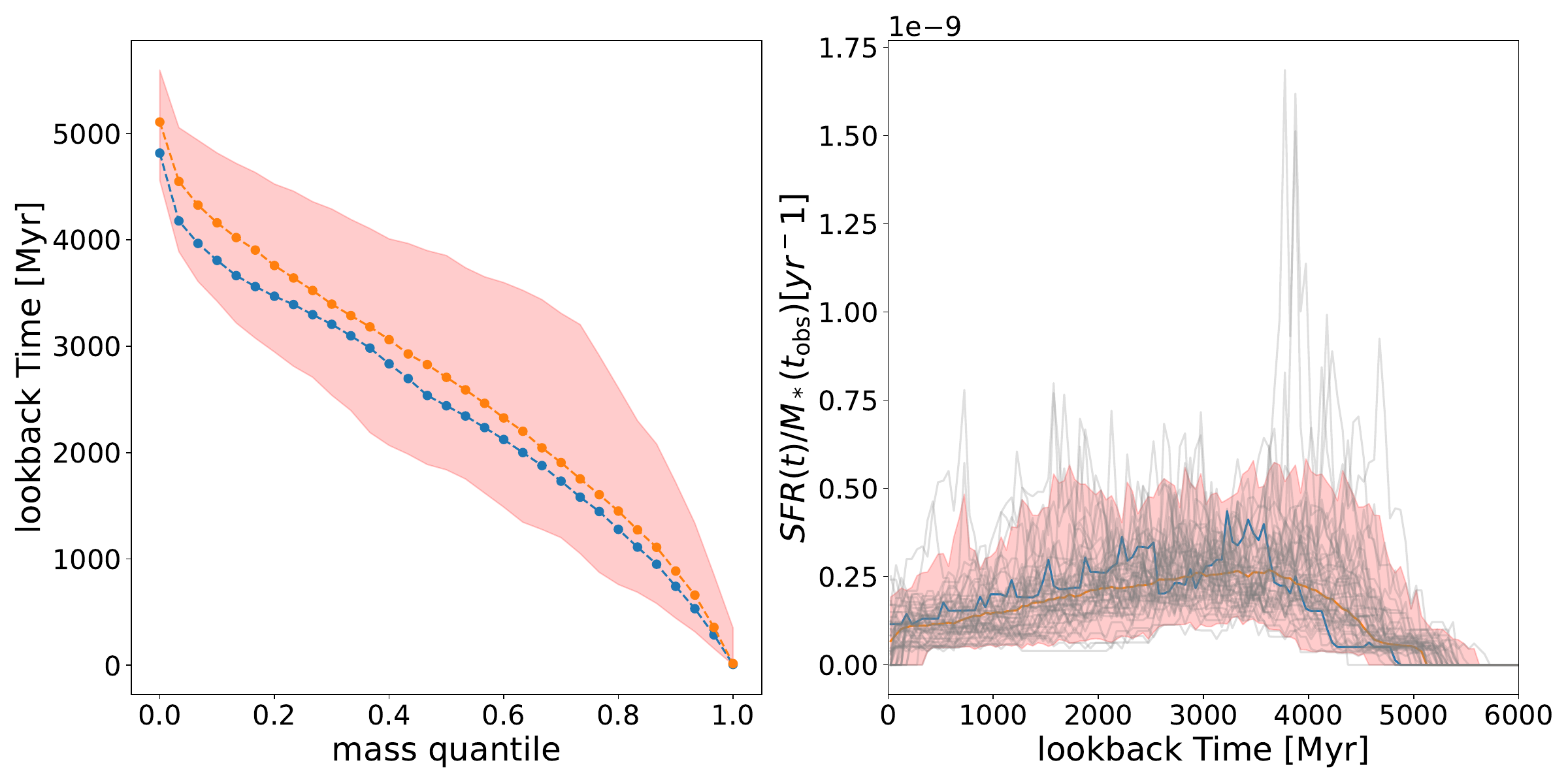}
 \caption{Example of the quantities estimated by our neural network {for a randomly selected galaxy in {\sc Horizon-AGN}}:  the true SFH is in blue, the estimated mean in orange and the 95\% confidence interval is in red. On the left the mass quantiles estimated by the neural network and on the right the resulting reconstruction. SFHs in grey are draws from the estimated posterior distribution.}
 \label{fig:example}
\end{figure}

\subsection{Parameterising star formation histories}
\label{sec:sfh}
To successfully reconstruct the SFH, it is essential to select an appropriate model that can be effectively inferred. Galaxy SFHs are inherently complex as they encompass a wide range of timescales, from the rapid bursts of star formation seen in starburst galaxies to the slow, steady star formation in more quiescent galaxies. Additionally, galaxies may experience episodic bursts of star formation due to external triggers, such as galaxy-galaxy interactions or mergers. These diverse scenarios give rise to a rich array of SFHs shapes. The challenge arises from the need to distil this complexity into a concise set of parameters that can be effectively learnt and predicted by neural networks, especially when we are interested in recovering the complete posterior distribution and not simply a maximum likelihood estimate.

Multiple parameterizations have been proposed, from the simplest constant SFR to stochastic processes \citep[e.g.][]{tacchella20, ciesla24}. Analytical models such as exponentially declining or \enquote{delayed} models \citep[e.g.][]{ ciesla17} also have extensively been explored \citep[e.g.][]{Ciesla18, Carnall19}.

We use here a formalism similar to that in previous works \citep{Iyer19, 2016ApJ...824...45P, 2019MNRAS.488.3143B}. 
The SFH is characterised using the total stellar mass $M_*$, with $N$ distinct lookback times corresponding to specific quantiles of $M_*$, and the initial age marking the onset of the SFH. For example, by selecting $N = 10$, it becomes necessary to determine the initial lookback time $t_0$ marking the start of the SFH, along with times $t_{10}$, $t_{20}$, ... ,   to  $t_{100}$ , which represent the moments when $10\%$, $20\%$, ... ,  to  $100\%$ of the galaxy final stellar mass has been formed, respectively. This latter time should be $t=0$ (in lookback time) for galaxies which are still forming stars, but could be different for completely quenched galaxies (Fig. \ref{fig:example}).

Increasing the value of $N$ leads to a more detailed representation of the SFHs \citep{Iyer19}. However, this improvement is accompanied by an increase in complexity in the estimation process as a consequence of the increase in the dimensionality of the parameter space. We set $N=30$ to allow for sufficient flexibility without being intractable. Figure \ref{fig:binning_hz} illustrates the the effect of this approximation.

\begin{figure*}
 \includegraphics[width=\columnwidth]{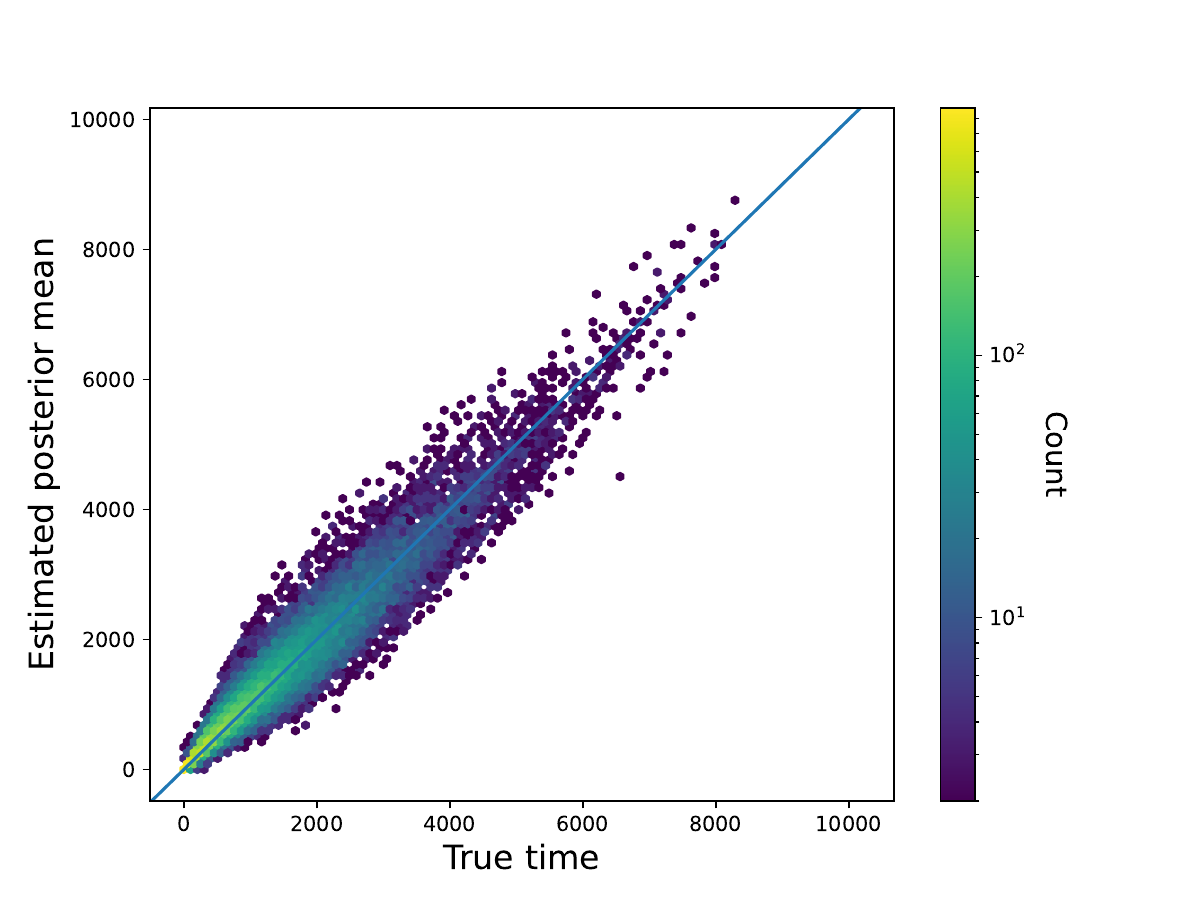}
 \includegraphics[width=\columnwidth]{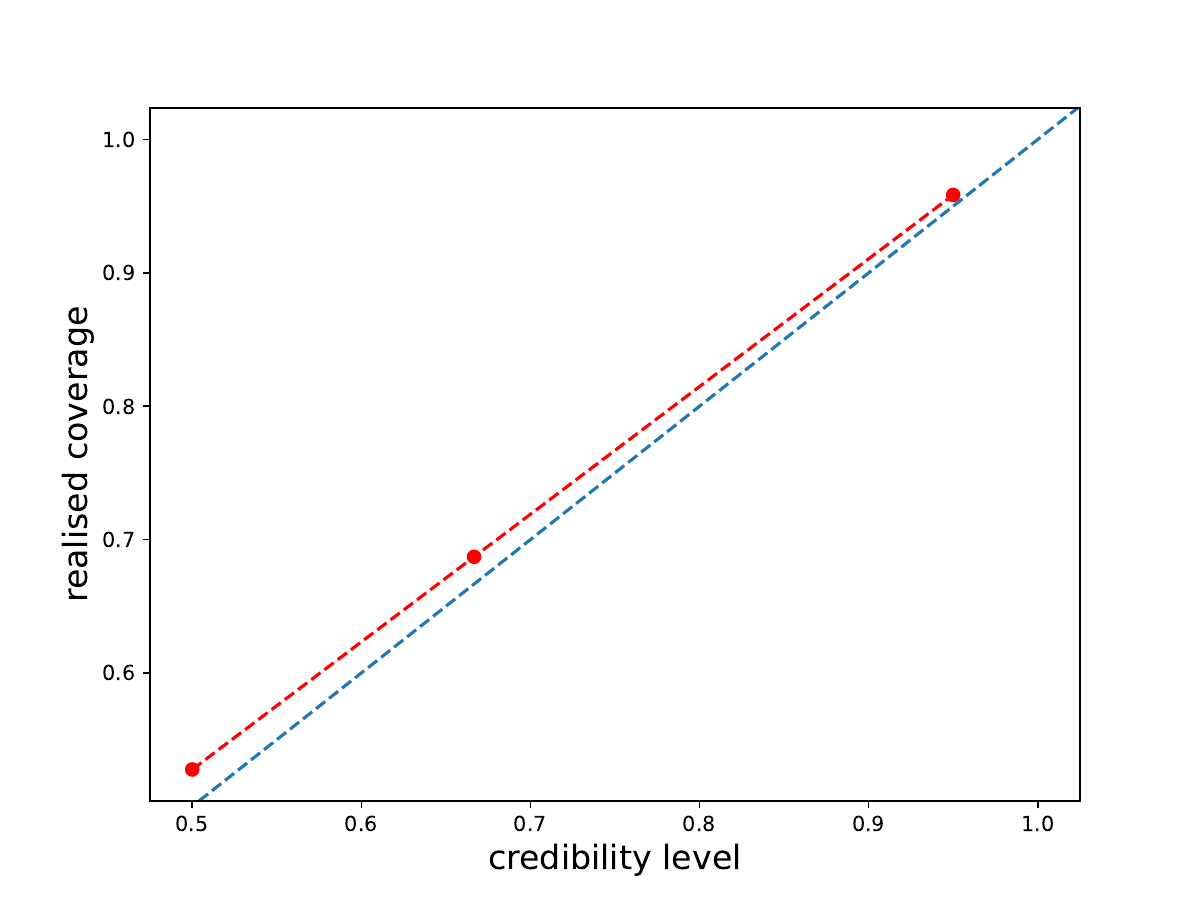}
 \caption{{Left} : Comparison between the true formation times and the estimated posterior mean in {\sc Horizon-AGN}. Perfect inference would lie on the (blue) diagonal line. Our methodology yields an unbiased estimator with regard to {\sc Horizon-AGN}, for all timescales. 
 {Right:} Assessment of the estimated uncertainty calibration with the credibility level on the x-axis, and  the realised coverage on the y-axis. For each level ($50\%,68\%$, and $95$\% ), the true value lied in the estimated credibility interval slightly more often than the expected level. Perfect uncertainty calibration would be on the blue diagonal. Our estimator is well calibrated. }
 \label{fig:perf_SBI}
\end{figure*}

\subsection{Simulation Based Inference}
\label{subsec:SBI}
In our technique, we use simulation-based inference to robustly estimate the SFHs of observed galaxies and provide proper Bayesian uncertainty measurements. This approach combines the power of cosmological simulations with deep learning to bridge the gap between numerical models of galaxy evolution and observed data.
An important advantage of cosmological simulations lies in their capacity to create synthetic populations of galaxies, each with a precisely known SFH. These simulations attempt to emulate the complexities of the observable Universe, generating synthetic galaxy catalogues that contain comprehensive details of the SFH for each galaxy. Such synthetic galaxies provide a means to directly correlate actual SFHs with the observed photometric data.

A significant advantage of Bayesian inference is its capacity to propagate uncertainty from prior knowledge and data into parameter estimates. These uncertainties are essential for assessing the reliability of our parameter estimates and provide a quantifiable measure of confidence in the inferred SFHs. The prior represents our initial beliefs or knowledge about the parameter before observing any new data, while the likelihood is a function that describes the probability of the observed data given a particular parameter value. 

Simulation-based inference relies on implicit construction of the likelihood function using samples drawn from a generative model. In the context of our SFH estimation, the likelihood function is constructed by associating the simulated photometry of a galaxy to its SFH.

In practice, SBI uses a deep neural network to parameterise a probability distribution \citep{2020PNAS..11730055C, JMLR:v22:19-1028}. This neural network is then optimised to minimise the Kullback-Leibler divergence \citep{NIPS2016_6aca9700} from the true posterior distribution implied by the generative model sampling the training data and the neural network output. This allows us to construct an estimator of the posterior density directly using a training set of pairs $(x_i, \theta_i)$, where $x_i$ are the observed quantities (here the photometry) and $\theta_i$ the corresponding parameters (here the formation times).  We refer the reader to \cite{NIPS2016_6aca9700,  2019arXiv190507488G}, and \cite{2024arXiv240205137H} for more details.
This work use the SNPE-C \citep{2019arXiv190507488G} implementation of the \texttt{sbi} package \citep{tejero-cantero2020sbi}, using a mask autoregressive flow \citep{papamakarios2018masked} with its  default ``sbi'' settings as the density estimator. The neural network is trained to estimate $p(t_0,\dots , t_{100} | x_c)$ the posterior distribution of the cumulative star formation times $t_0,\dots , t_{100}$ (see Section \ref{sec:sfh} ) given a vector of colours $x_c$. The only observed quantities used for inference and model checking (see Section \ref{sec:outlierdetection}) are the colours corresponding to consecutive bands:  $u^*-g, g-r, r-i, i-z, z-Y, Y-J$, $J-H$ and $H-K_\mathrm{s}$. 

In order to standardise the problem, the times $t_0,\dots , t_{100}$ are re-scaled by a constant $T$ such that $T>t_0$ for all galaxies in the training set. $T$ is the same for all galaxies in the sample, and slightly higher than the largest value in the training sample. We therefore have 
\begin{equation}
0\leq t_i<1 \quad \forall i \in [0,100]
 \end{equation}
and we rescale back $t_i$  after inference is complete. This restricts the estimated posterior support, allowing Monte-Carlo sampling via the accept-reject algorithm using a uniform distribution over the unit hypercube. As is common with neural network applications for numerical stability \citep{lecun-gradientbased-learning-applied-1998}, inputs are normalised by removing the mean and dividing by the standard deviation of the colours in the training set. 

Figure~\ref{fig:perf_SBI} shows the performance of our estimator on a test set of $\sim 85000$ randomy-sampled Horizon-AGN that were not used during training. The left panel compares the true times $t_i$ associated with the mass quantiles as defined above to the mean of the estimated posterior distributions and shows very good agreement, with a reasonable scatter and no significant bias. The right panel shows calibration of the credible intervals: for each galaxy in the test set, we estimate the $50\%,68\%$, and $95$\% credible intervals, and check if the true value lies in each interval. A more detailed assessment is presented in Appendix~\ref{fig:quantiles}, where we show the error of the formation times of each mass decile. Once again the estimates are always unbiased, with a scatter $\sigma_{\rm err}\leq 0.16$\,dex. We then compute the proportion of intervals that pass this check. For a perfectly calibrated estimator, the $50\%$ (resp. $68\%$, and $95$\%) intervals should contain the true value  $50\%$ (resp. $68\%$, and $95$\%) of the time. Our estimator yields well calibrated intervals.

\begin{table}
{\small
\centering
\caption{Algorithms used for outlier detection in order to perform prior checking}
\begin{tabular}{ |c||c| }
 \hline
 Algorithm & Reference\\
 \hline
 Histogram-base Outlier Detection    & \cite{hbos}\\
Isolation Forest    & \cite{IF}\\
K-Nearest Neighbours (KNN)  & \cite{10.1145/335191.335437}\\
Local Outlier Factor    &  \cite{LOF}\\
Minimum Covariance Determinant  & \cite{MCD}\\
Principal Component Analysis   &\cite{Aggarwal}\\
Gaussian Mixture Model & \cite{2019AnRSA...6..355M}\\

 \hline
\end{tabular}
\label{table:outlier_algos}
}
\end{table}

\begin{figure}
 \includegraphics[width=\columnwidth]{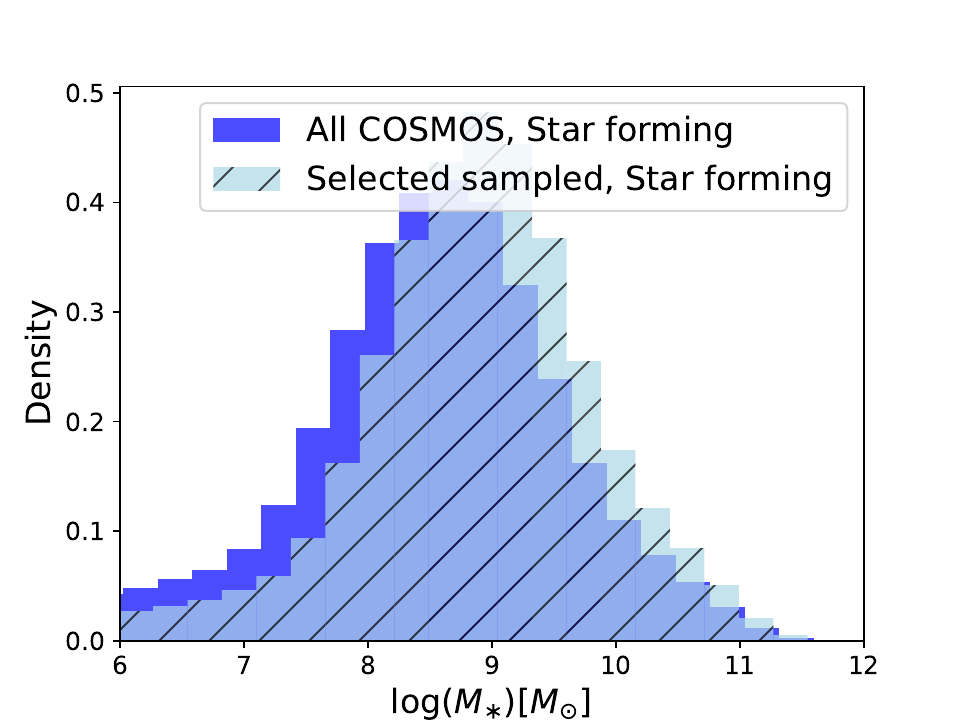}
  \includegraphics[width=\columnwidth]{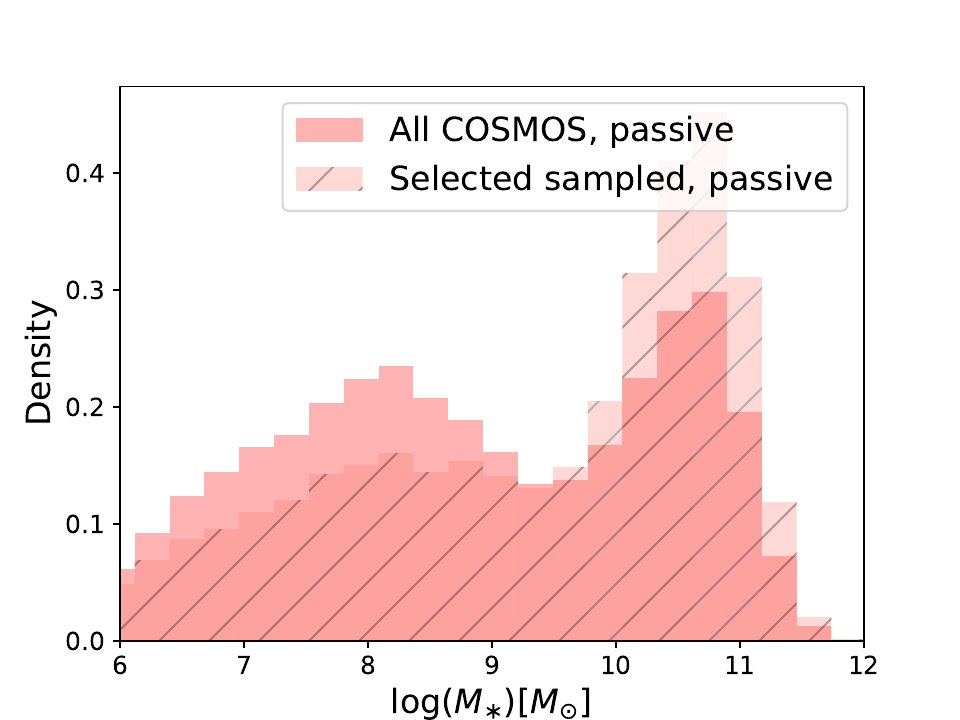}
 \caption{Normalized mass distributions of galaxies in our COSMOS2020 sample. {Top}: Star-forming galaxies. {Bottom}: passive galaxies. The mass distribution of all galaxies in COSMOS (meeting requirements of observed bands, depth and SNR, see Section \ref{sec:data}) is darker and highlights potential biases in our selected sample after outlier rejection (lighter). }
 \label{fig:popoutliers}
\end{figure}

\begin{figure}
 \includegraphics[width=\columnwidth]{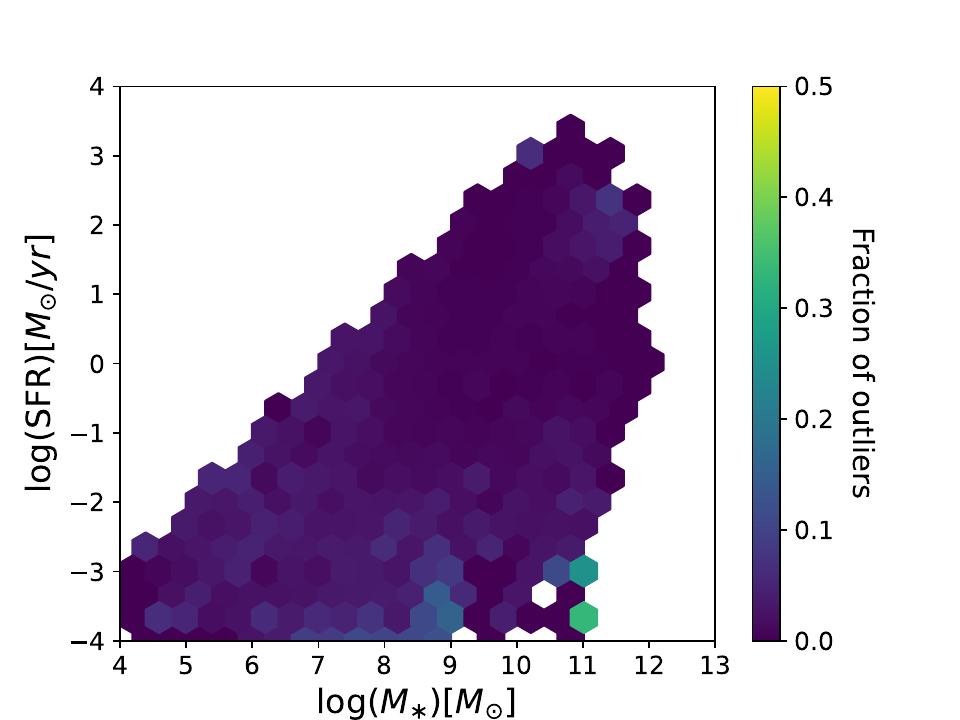}
 
  \includegraphics[width=\columnwidth]{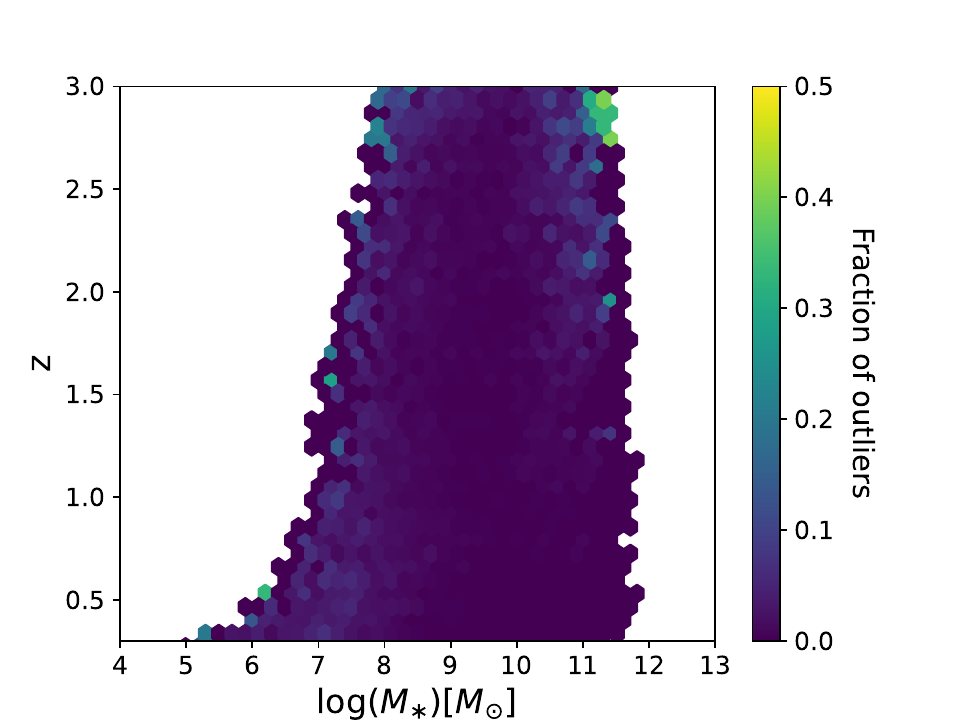}

 \caption{Star-formation rate and redshift of galaxies in the COSMOS2020 catalogue that are rejected by our prior predictive check, as a function of their mass. Galaxies that are classified as outliers (see Section~\ref{sec:outlierdetection}) are excluded from our sample for being too different in the colour space from the galaxies represented in {\sc Horizon-AGN}.}
 \label{fig:popoutliers2}
\end{figure}

\subsection{Prior checking, outlier detection and data selection}
\label{sec:outlierdetection}
As with every supervised learning approach, SBI requires that the training data match the inference dataset closely. From a computational perspective, this ensures that the algorithm is optimised on a suitable  region of the data space, and from a Bayesian inference perspective this is equivalent to prior checking \citep[e.g.][]{gelmanbda04,10.1214/06-BA117A}. More specifically we need to ensure that the photometry simulated from {\sc Horizon-AGN} is able  to adequately model the observed galaxies. If the {\sc Horizon-AGN} simulation cannot produce photometry that resembles the observations, we cannot meaningfully perform inference using it in our prior assumptions.

Therefore, we need a way to check, for every observed galaxy, if the simulation is able to produce galaxies with photometry similar to observations. The Bayesian literature refers to these assessments as `prior predictive checking' or `posterior predictive checks' \citep[e.g.][]{62bfc978-09b1-3997-9776-380d0b45e9c2}. Prior predictive checking consists of simulating mock data from the statistical model by sampling the prior distribution and assessing through a statistical test whether the observation is well covered by the resulting (so called ‘‘prior predictive’’) distribution. If the observation is too far in the tails of the prior predictive distribution, the model is considered ill-suited to perform inference. Posterior predictive checking describes a similar procedure, but replaces sampling from the prior distribution by sampling from the posterior distribution obtained after fitting the data first. This allows us to reduce the importance of the prior distribution in the check but is too computationally expensive for our use case, as it requires being able to fit each observation and make new simulations according to this fit for each observed galaxy.

To perform prior predictive checking, we approach it as a problem of outlier detection. The concept of outlier (or anomaly) detection involves establishing a reference distribution and identifying observations that significantly deviate from this norm. By treating the prior predictive distribution as this `reference', using outlier detection becomes an effective method for evaluating how well the simulations align with actual observations.

Multiple algorithms spanning the entire machine learning literature have been proposed to perform outlier detection, from direct density estimation using parametric methods (such as Gaussian Mixture Models, GMM, \citealp{2019AnRSA...6..355M}) or non-parametric approaches (such as histogram-based estimators \citealp{histogram,hbos}) to deep learning or kernel-based approaches \citep[e.g.][]{10.1145/3394486.3403062, 9347460}. We aggregate several different algorithms to assess whether an observation is accounted for or not in the simulation. 

In fact, aggregating different outlier detection algorithms is essential to mitigate the biases introduced by the specificity of a given algorithm, as it addresses several critical needs in outlier detection and model checking. By combining multiple algorithms, we reduce the risk of relying on a single algorithm that may be prone to producing biased results. This robustness improves the reliability of outlier detection. Different algorithms have different underlying assumptions and approaches for identifying outliers. Aggregating them provides a more comprehensive view of what constitutes an outlier in the data. This diversity of perspectives can help capture outliers that may be overlooked by individual algorithms. 
In practice we use the \textit{pyod} Python package \citep{zhao2019pyod} which contains the implementation of many outlier detection algorithms. From these, we select seven that are (see Table  \ref{table:outlier_algos}) well-suited to our dataset in terms of dimensionality, number of galaxies and computational cost. Each algorithm produces a so-called `outlier score' assessing the compatibility of a given data point with the estimated `reference distribution'. For example, in the GMM case, we fit a GMM to the photometric training set. The outlier score for each observed photometry is directly given by the value of the GMM density evaluated at that point. Observations far from the training set will have low density and, therefore, a low score. By setting a threshold, we can categorise which points are deemed outliers according to the scoring of the algorithm. The selection of this threshold is analogous to the construction of a confidence interval; it is set to reject a specified percentage of the reference distribution as outliers.

For each algorithm, the score threshold is chosen so as to reject 1\% of the simulations as outliers. Each observed galaxy is fed to each of the seven algorithms, and a majority vote among the algorithms dictates whether to reject it or not. A galaxy needs to be classified as an outlier by at least four algorithms to be rejected. This ensures that our detection process is robust to individual algorithmic biases.

\subsection{Application of the outlier rejection method to COSMOS2020}
\label{sec:outlierdetectioncos}

We apply the prior checking procedure described above to the COSMOS2020 catalogue. If COSMOS2020 and {\sc Horizon-AGN} photometric catalogues followed exactly the same distribution about $1\%$ of galaxies in COSMOS2020 would be rejected. Remarkably, less than $2\%$ of our catalogue is rejected despite the known differences in the physical parameters between the simulation and the observed galaxies. This means that while physical properties differ, the span of the colour distribution is similar in the simulation and in the observed catalogue. We limit our study to those observed galaxies whose photometric properties are adequately represented in the simulation, regardless of physical properties. {Figure~\ref{fig:popoutliers2} presents the fraction of outliers as a function of SFR, mass and redshift. The fraction is slightly higher for massive galaxies, especially at high redshift and with very low or very high SFR. This indicates that the colours of these populations are not perfectly reproduced in the {\sc Horizon-AGN} simulation. 
Interestingly, the fraction of outliers stays low for observed galaxies with $\log M_*/M_\odot < 9$, although the simulated sample contains only galaxies with $\log M_*/M_\odot > 9$. This means that these low-mass galaxies have a colour distribution similar to those of more massive galaxies in the simulated sample, despite their different masses.}

\label{subs:mass_type}
\begin{figure}

 \includegraphics[width=0.95\columnwidth]{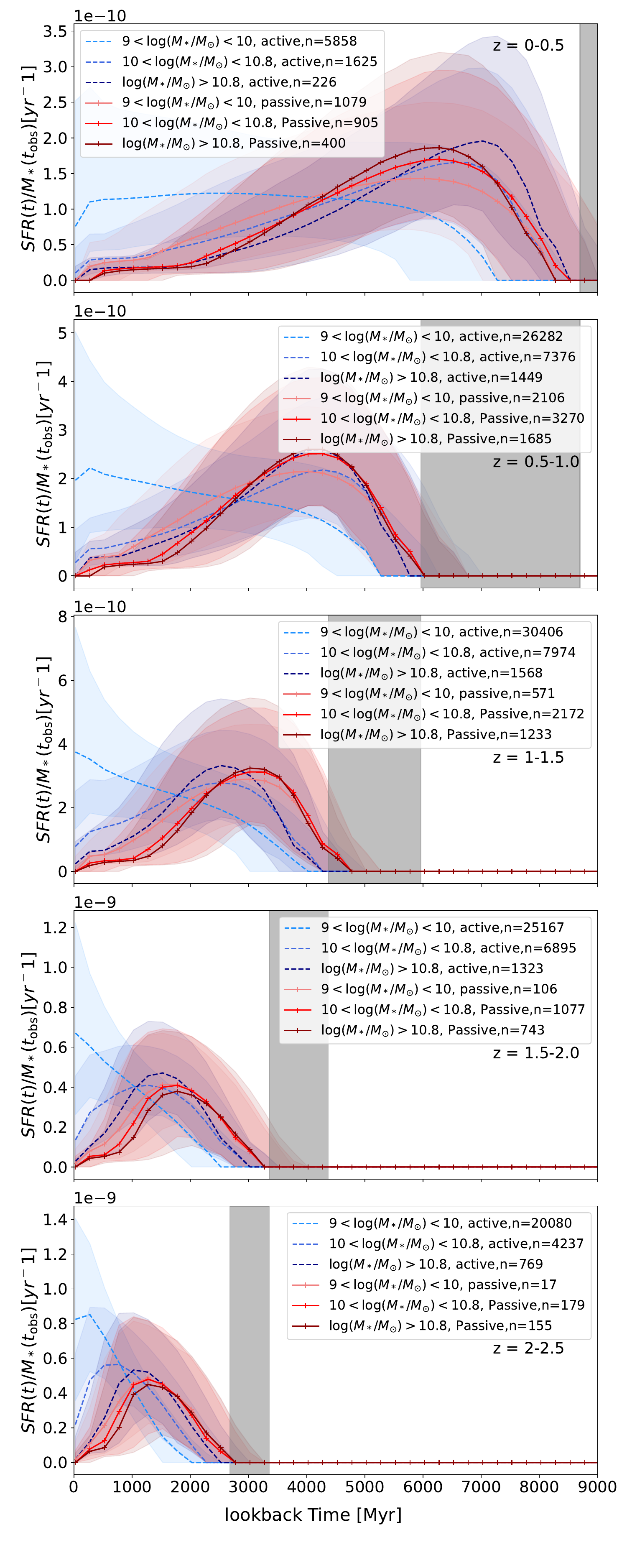}

 \caption{Median SFHs of star-forming (blue dashed lines) and passive (red sold lines) galaxies in COSMOS2020, in several redshift ranges and decomposed in three mass bins as indicated in the panels. The number of galaxies used to compute the median trend is indicated in the legend. The vertical grey shaded area (the width of which depends on the considered redshift range) indicates the maximum lookback time, above which the start of the SFH is not compatible  with the age of the Universe.}  \label{fig:activepassive}
\end{figure}

\begin{figure}
 \includegraphics[width=1.05\columnwidth]{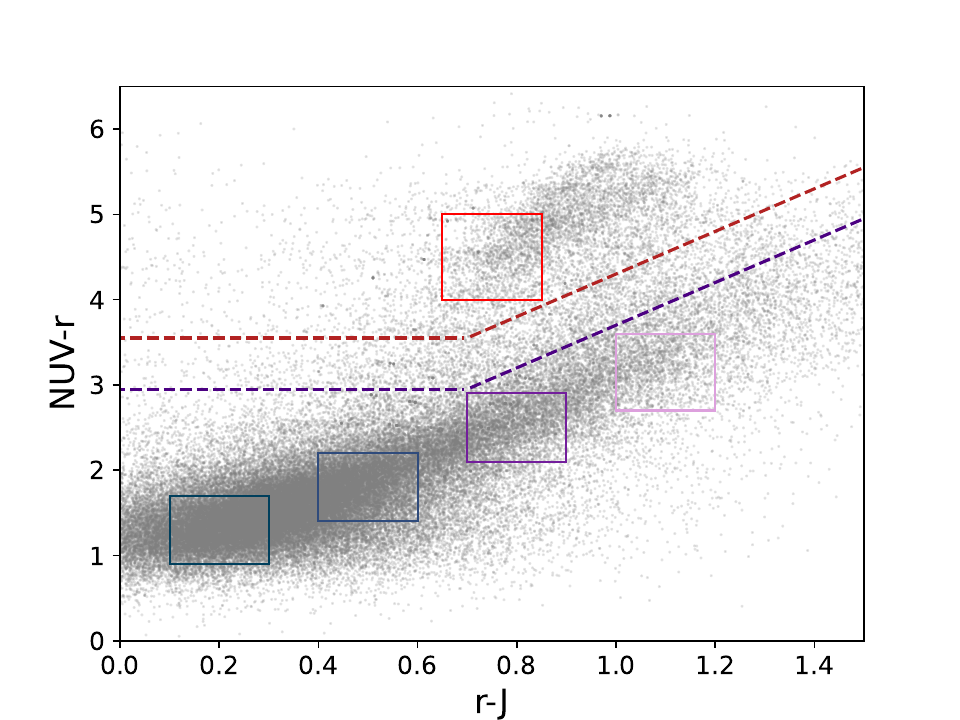}
  \includegraphics[width=0.99\columnwidth]{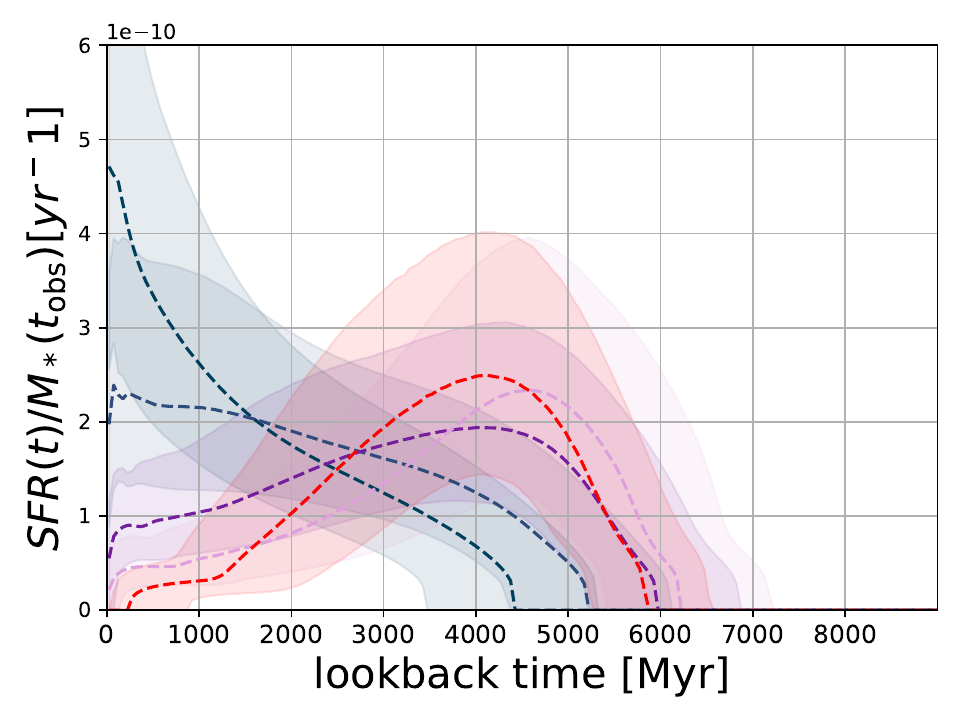}
 \caption{Evolution of the estimated SFHs for galaxies at redshift $0.5<z<1$ in COSMOS2020, as a function of their position in the $\rm{NUV}\mathit{rJ}$ diagram. {Top}: four galaxy populations are selected in the star-forming region (blue and clear boxes) and in the passive region (red box).  {Bottom} : Estimated SFHs of galaxies in each box (same colours). The SFHs vary smoothly from steeply increasing and younger star formation to older SFHs resembling the analytical `delayed' model, going from left to right in the diagram. The SFHs reconstructed for galaxies in the passive region of the $\rm{NUV}\mathit{rJ}$ diagram are quenched, as expected.}
 \label{fig:NUVrJ_patches}
\end{figure}

\section{Results}
\label{sec:results}

As illustrated in Fig.~\ref{fig:example}, our method, when considering the median of the posterior distribution, is ideal to understand the general build-up of stellar mass in galaxies, but it is less sensitive to short time-scale processes or stochasticity such as multiple close bursts. We caution that the outlier rejection performed in Section~\ref{sec:outlierdetectioncos} implies that our analysis sample may be incomplete and potentially biased toward {\sc Horizon-AGN}. Fig.~\ref{fig:popoutliers} illustrates how the various cuts impact the mass distribution of star-forming and passive galaxies. This will be further discussed in Section~\ref{subsec:validity}. Nevertheless, we also note that the training is performed on colours instead of fluxes and magnitudes. Therefore, any dependence on the mass of the SFH is a result of the model's prediction, not directly derived from the mass-luminosity relation. This is also the reason why we choose not to apply a mass limit, even if the simulated catalogue used in the training is mass-limited.

\subsection{Median SFHs of galaxy populations}
\label{sec:medianSFHs}

Figure~\ref{fig:activepassive} shows the median SFH of passive and star-forming galaxies in COSMOS2020, in different mass bins as a function of lookback time over the redshift ranges $0.5<z<1$, $1<z<1.5$, $1.5<z<2$ and $2<z<2.5$ (the number of passive galaxies in the $2.5<z<3$ bin was too low to be analysed).

To determine these trends, we proceed as follows. For each galaxy, we draw $50$ SFHs from the posterior distribution. These are grouped depending on mass, photometric redshifts and type (star-forming or passive). The figure shows the median and $68\%$ interval of each group. As a reminder, our SFH reconstruction approach does not require galaxy types in input. Types are inferred independently from the rest-frame ${\rm NUV}-r$ versus $r-J$ diagram (hereafter $\rm{NUV}\mathit{rJ}$) colour diagram. For this reason, checking that the median SFH for a given galaxy type is in broad agreement with expectation is a validation test. Indeed we visually note that red galaxies have little to zero recent star formation in their reconstructed SFHs, while blue galaxies generally have much higher recent star formation. On overall recent star-formation is higher in lower masses galaxies, {a manifestation of the downsizing phenomena, i.e. the fact that massive galaxies have formed most of their mass more rapidly and earlier in the history of the Universe \citep[see e.g.][]{2006MNRAS.372..933N}. } 

We also note that at all redshifts there is very little evolution of the median SFHs of passive galaxies when divided by mass (while the median SFH of star-forming galaxies is strongly mass-dependent as described above). However, for both populations  the 68\% intervals are very extended, suggesting a strong diversity of SFHs within a mass range. 

This diversity is expected to manifest itself in the formation time, the overall slope of the SFH after the peak of star formation and the smoothness of the SFH (or number of local maxima). These quantities will be further examined in Section~\ref{sec:slope}.

We next consider the behaviour of the reconstructed SFHs for different galaxies in the ${\rm NUV}rJ$ diagram. This technique \citep{ilbert13} is now a well-tested method to separate star-forming and passive galaxies using a rest-frame two-colour selection.  Fig.~\ref{fig:NUVrJ_patches} shows that the reconstructed SFHs vary smoothly across the ${\rm NUV}rJ$ diagram from young and very star-forming galaxies on the bottom left (dark blue curve) to passive galaxies on the top (red curve). We observe no abrupt change in SFH between galaxies close in the rest-frame colour space, and the recovered SFH shapes are in agreement with \cite{ilbert13}. However, the shape of the SFH of massive galaxies classified as star-forming (in the ${\rm NUV}rJ$ diagram) is questionable (light purple curve), as it seems to fall to zero at the time the galaxies are observed. The possible causes for this will be discussed in Section~\ref{subsec:dust_problem}.

In addition to showing consistent type-dependent shapes, the reconstructed SFHs are also consistent in terms of galaxy ages. The grey-shaded region shows the maximum lookback time beyond which the SFH would be incompatible with the age of the Universe, varying with redshift. For example, at $0.5<z<1$, the maximum lookback time of the most massive galaxies is~$\sim 6\pm 0.5$\,Gyr, which suggests a formation time at most~$\sim 13.8 \pm 0.5$\,Gyr ago, consistent with the age of the Universe within the photometric redshift uncertainties. Similarly in other redshift bins, the formation times of galaxies are compatible with the age of the Universe if we apply to the maximum lookback time the time shift deduced from the photometric redshift of the galaxies. This result is particularly remarkable given that the galaxy photometric redshifts were not used as input to our reconstruction algorithm.

\begin{figure*}
\begin{center}
      \includegraphics[width=0.9\textwidth]{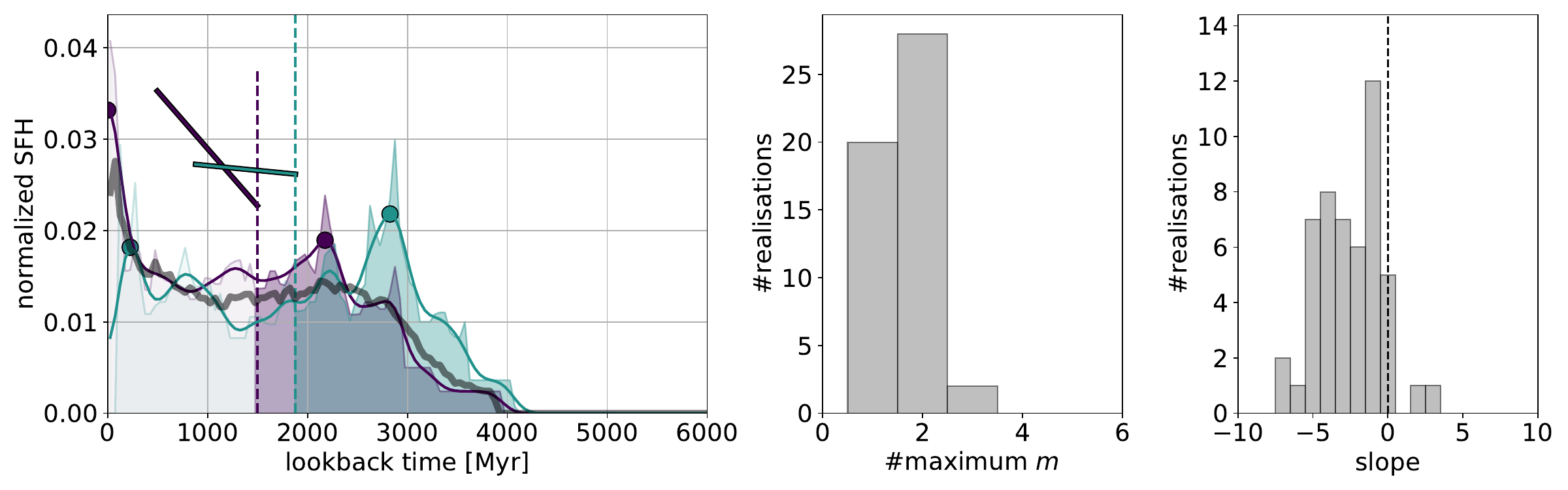}
 \caption{Illustration of our method to measure the formation age $t_{\rm form}$, slope, number of maxima and age of the first peak for an individual SFH drawn from the posterior distribution of each galaxy. On the left panel are two realisations of the SFH for the same galaxy from COSMOS2020 catalogue (purple and cyan lines, smoothed with a Gaussian kernel of standard deviation 100~Myr), together with the median over 50 realisations (grey line). The dashed vertical line indicates $t_{\rm form}$, which is then used to compute the slope (solid straight segments). The circular points indicate the position of the persistent maxima. The median and right panels show the histograms of the numbers of maxima and  slopes  over 50 realisations.}
  \label{fig:method_summary}
 \end{center}
 \end{figure*}

\begin{figure*}
\begin{center}
  \includegraphics[width=0.9\textwidth]{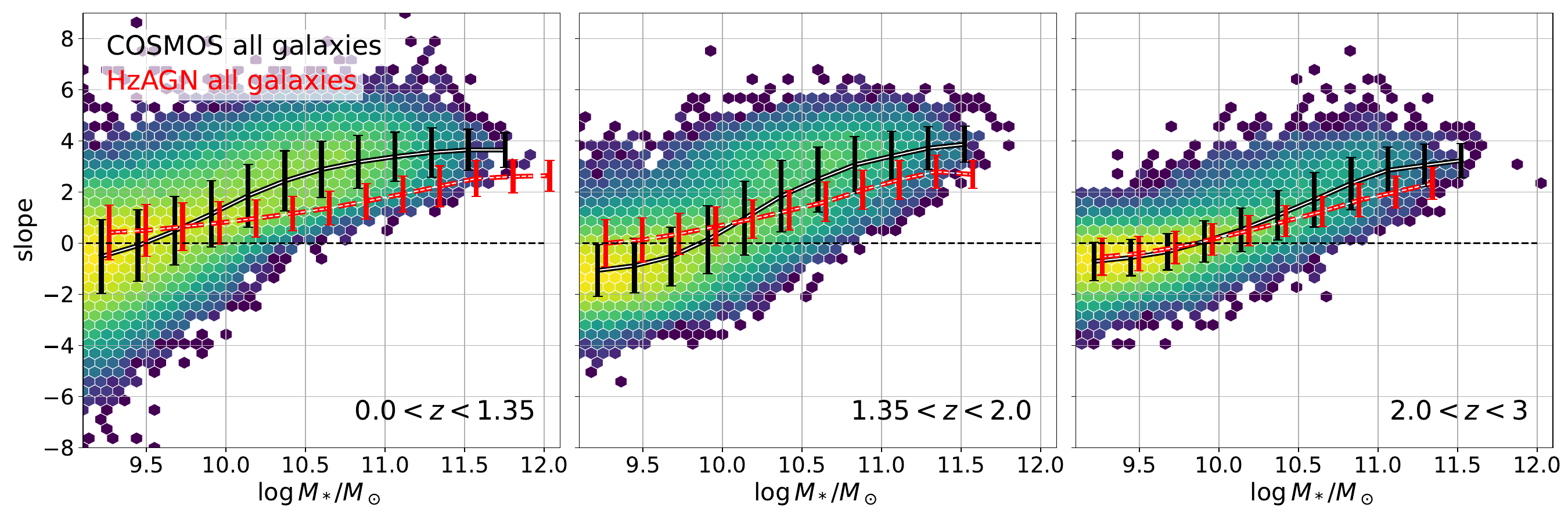}
 \includegraphics[width=.9\textwidth]{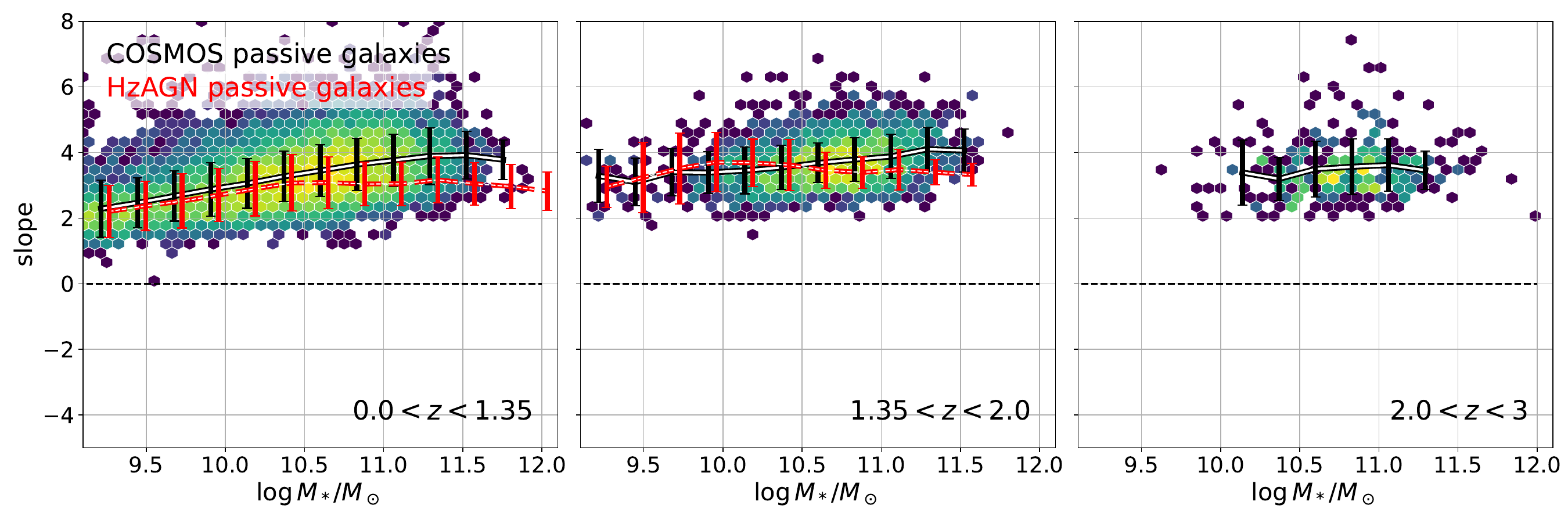}
\end{center}

 \caption{ Slope of the reconstructed SFH, as defined in equation~\ref{eq:slope} in COSMOS2020 (black solid lines with errorbars) and {\sc Horizon-AGN} (red dashed lines) for all galaxies in the sample ({top}) and passive only ({bottom}). For each galaxy, individual slopes computed from 50 SFHs drawn from the posterior distribution are averaged. The two first redshift ranges have been chosen to encompass similar comoving volumes in order to mitigate possible cosmic variance effects in the galaxy distribution. The hexagons are colour-coded by the logarithm of the galaxy number density in COSMOS2020 in the considered redshift bin. }
 \label{fig:slopecos}
\end{figure*}
\subsection{Characterisation of individual SFHs and formation redshifts}
\label{sec:slope}

The nature of processes responsible for shaping star formation and quenching is still unclear and is expected to vary depending on the mass and the environment of the galaxies. For example, quenching could be associated with a cut-off of the gas supply, leading to a slow cessation of star formation as the cold gas supply runs out. This cutoff could be driven by tidal effects or stripping processes known as strangulation \citep{balogh20,Bosch08} in the case of low-mass galaxies in dense environments, or gas heating via shocks \citep{Dekel2006} or AGN \citep{Dubois2013} in the case of massive galaxies. Depending on the nature of the process, the quenching timescales could be different. Quenching timescales due to AGN are unclear, as they can also eject strong outflows, which immediately remove the gas from the galaxies and hence turn off the star formation \citep{Springel2005}. Violent events such as disk instabilities or mergers \citep{snyder11} can also temporarily lead to a very rapid exhaustion of the gas reservoir via an intense burst of star formation, but in turn rapidly quench the galaxy afterward. For galaxies in the hot intracluster medium, gas can be quickly removed via ram-pressure stripping \citep{abadi99}, which finally halts star formation on a short timescale \citep[see also][]{park22}. 

Although this list is not exhaustive, we emphasise that given their different timescales, these processes will have different impacts on the shape of the SFH. Therefore, it is essential to derive relevant summary statistics to identify the characteristic patterns and their variation with mass and redshift.

From each SFH we measure the following summary statistics, described below: the slope, the formation time (i.e. the age of the Universe when the galaxy was aged its mean age)  $t_{\rm f}$ and redshift $z_{\rm f}$, the number of maxima $m$ and the time of the first peak.
\subsubsection{Mass-weighted formation time $t_{\rm f}$} 
The mass-weighted formation age $t_{\rm form}$ is defined as
\begin{equation}
\label{eq:tf}
t_{\rm form} =
\frac{\int_{0}^{t_{\rm first}}t {\rm SFR}(t) {\rm d}t}{\int_{0}^{t_{\rm first}}{\rm SFR}(t) {\rm d}t}\,,
\end{equation}
where $t_{\rm first}$ is the lookback age corresponding to the start of the SFH and $t=0$ corresponds to the observation. $t_{\rm form}$ is in unit of lookback time. In practice, it is almost equivalent to the lookback age at which the galaxy had formed half of its stellar mass ${\rm t}_{50}$. Using $t_{\rm form}$ and the photometric redshift in COSMOS2020 or the exact galaxy redshift in {\sc Horizon-AGN} we then define the formation time $t_{\rm f}$, which is the age of the Universe at $t_{\rm form}$, and the formation redshift $z_{\rm f}$. 

\subsubsection{Slope} 

The slope estimates the overall trend of the SFH since the formation time $t_{\rm form}$, and is defined as:
\begin{equation}
\label{eq:slope}
    {\rm slope} \left[{\rm dex}\right] = \frac{\log {\rm SFR}_{95} - \log {\rm SFR}_{{\rm form}}}{t_{95}-t_{\rm form}} \times t_H(z)\,,
\end{equation}
where ${\rm SFR}_{95}$ is the SFR at the time $t_{95}$ where the galaxy has formed 95\% of its mass, ${\rm SFR}_{{\rm form}}$ is the SFR at the time $t_{\rm form}$ and $t_H(z)$ is the Hubble time at $z$. The slope is measured on a smoothed SFH, where the SFH is smoothed with a Gaussian kernel of standard deviation 100~Myr, as shown on Fig.~\ref{fig:method_summary} (compare the thin and thick purple and cyan lines on the left panel). We choose $t_{95}$ because it is the last estimated time by our algorithm before $t_{100}$, and this avoids having undefined slope values if ${\rm SFR}_{100}=0$, whereas having  ${\rm SFR}_{95}=0$ is less likely given the smoothing.
A positive slope indicates that the galaxy forms on overall less stars now than it was forming at $t_{\rm f}$. 

\subsubsection{Number of maxima and age of the first peak} 
\label{sec:nmaxima}
We also estimate the number of maxima $m$. Local maxima are identified and are filtered to keep only the highly persistent ones\footnote{We use this freely distributed implementation: \href{https://www.csc.kth.se/~weinkauf/notes/persistence1d.html}{https://www.csc.kth.se/$\sim$weinkauf/notes/persistence1d.html}}. The persistence of a peak simply quantifies the longevity of this peak as an isolated component in an excursion set (the set of values greater than a threshold) evolving with a decreasing threshold \citep[see e.g.][]{Edelsbrunner02}. The most persistent peaks are therefore the most prominent ones in the normalised SFH. The number of maxima is therefore dependent on the chosen persistence threshold, which should be chosen in the same unit as the SFH. Because all SFHs sum to 1, galaxies that formed a long time ago will have on overall smaller SFH values than those that formed recently. Therefore, we choose a persistence threshold of the same order of magnitude as the standard deviation of different realisations at a given timestep (each realisation is normalised by the number of non-null values in each SFH). We caution that this extraction of maxima is performed on the normalised SFH. Therefore, the \textit{absolute mass} formed under a SFH peak at a given persistence in high-mass galaxies will be higher than in low-mass galaxies. However, the fraction of the mass formed would be comparable. In this sense, the number of maxima allows one to quantify the smoothness of the SFH.

We note that varying this persistence threshold around this value does not change the relative trend within the galaxy populations but simply shifts it up or down. Finally, we also estimate the age of the Universe at the first peak of mass assembly $t_{\rm first\, peak}$.

For each galaxy, these quantities are estimated on 50 individual realisations of the SFH from the posterior and then averaged. Figure~\ref{fig:method_summary} illustrates the method of measuring these quantities in two individual realisations. The quality of these estimators is studied in Appendix~\ref{ap:qa}, relying on the comparison between the measurement of the true SFH in {\sc Horizon-AGN} and their reconstructed version.

\subsection{The diversity of SFH slopes as a function of mass}

Figure~\ref{fig:slopecos} shows the slopes of the reconstructed SFH as a function of stellar mass for the full population (top) and the quiescent population (bottom, identified from the ${\rm NUV}rJ$ diagram) in both COSMOS2020 and {\sc Horizon-AGN} respectively. The same SNR cut is applied to {\sc Horizon-AGN} and COSMOS2020. The error bars indicate the standard deviation around the mean. The redshift ranges have been chosen to encompass similar comoving volumes in COSMOS, in order to mitigate the possible effect of cosmic variance when comparing one redshift bin to the other. We note  however that the intrinsic 100 Mpc$/h$ size of the native {\sc Horizon-AGN} box from which the lightcone has been extracted implies for {\sc Horizon-AGN} a cosmic variance at 100~Mpc$/h$ scale. Slopes are estimated on 50 realisations drawn from the posterior distribution and then averaged. 

There is a large dispersion of slopes at a given mass, which is consistent with the extended 68\% interval in Fig.~\ref{fig:activepassive}. In addition, there is a clear dependency on mass for the entire population, and this dependency is larger in COSMOS2020 than in {\sc Horizon-AGN}. As expected, the slope is larger for higher mass galaxies, indicating that they formed most of their mass closer to the formation time ${\rm t}_{\rm form}$ and are more likely to have quenched by the time they are observed. On the other hand, low-mass galaxies have a negative or very small slope. We note that the low-to-high mass transition is more pronounced in COSMOS2020 than in {\sc Horizon-AGN}. In particular, observed galaxies form stars more vigorously compared to simulations, with negative slopes (indicating rising SFHs) up to $ \log M_*/M_\odot < 10.5$. This behaviour was previously highlighted by \cite{kaviraj17} who found that the low-mass end of the star-forming main sequence {formed stars less vigorously} in {\sc Horizon-AGN}. 

We find that massive galaxies generally have larger positive slopes in COSMOS2020 compared to {\sc Horizon-AGN}, indicating possibly more drastic quenching. This is in agreement with \cite{dubois16}, who showed that massive simulated galaxies generally contain larger residual star formation than in observations. Because we normalized the slope by the Hubble time $t_{H(z)}$, at a given mass, there is no redshift dependency of the median slope in COSMOS2020, but we find a larger dispersion at lower redshift.

For passive galaxies, the slope weakly depends on stellar mass in the two lowest redshift bin in COSMOS2020, with the SFHs of more massive galaxies dropping slightly more rapidly since their formation time than lower-mass galaxies. This trend is also consistent with Fig.~\ref{fig:activepassive}. This behaviour is not seen in {\sc Horizon-AGN} or in higher redshift bins. However, at a given stellar mass, we also note the large diversity of possible slopes, especially at low redshift, which could point to distinct quenching mechanisms. 
 
Refinement of the analysis of slope as a function of other quantities, such as environment or halo mass, would be a promising avenue to understand the quenching mechanisms of passive galaxies but is beyond the scope of this study.

\subsubsection{Number of maxima}
\label{sec:dissectionSFH}

\begin{figure}
\begin{center}
  \includegraphics[width=1.0\columnwidth]{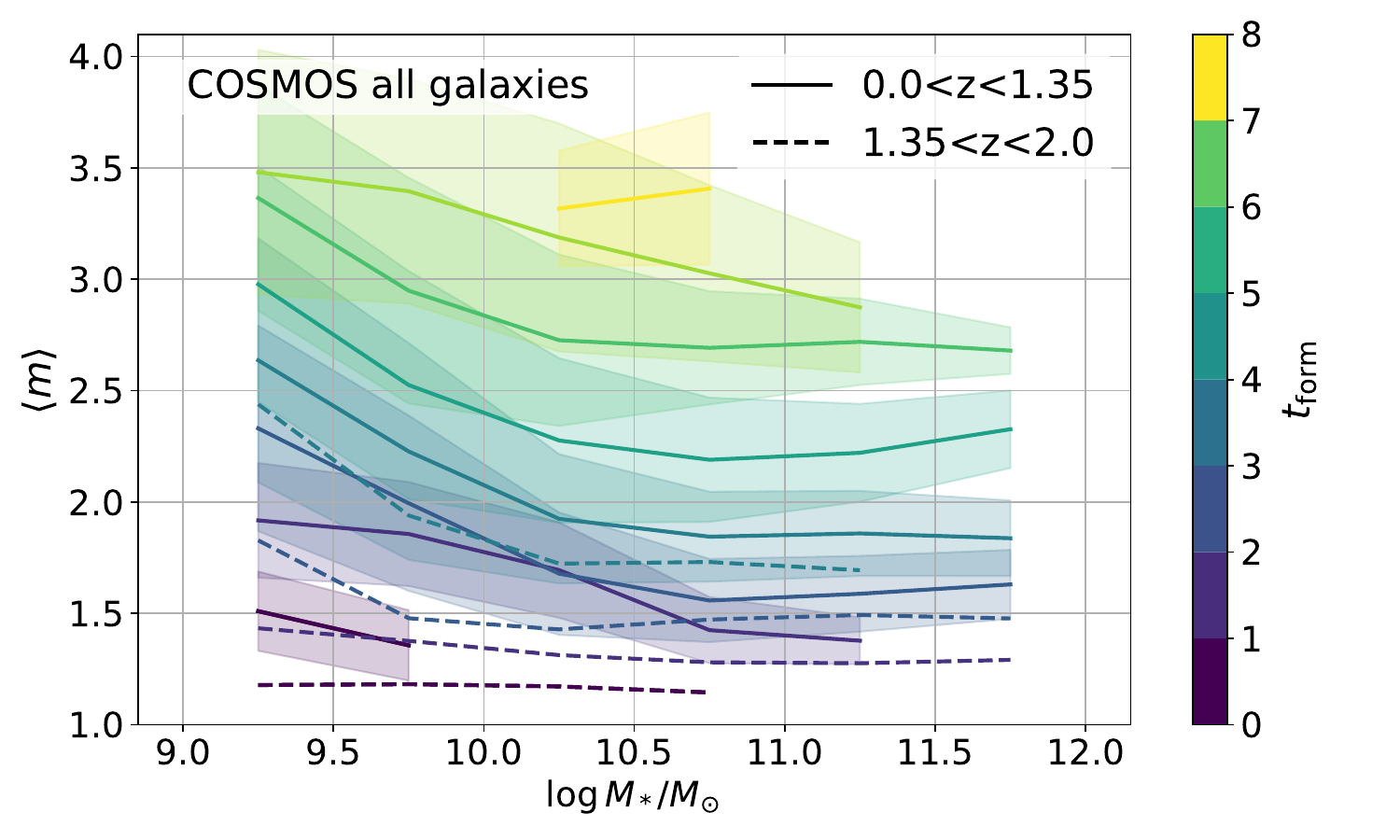}
  \includegraphics[width=1.0\columnwidth]{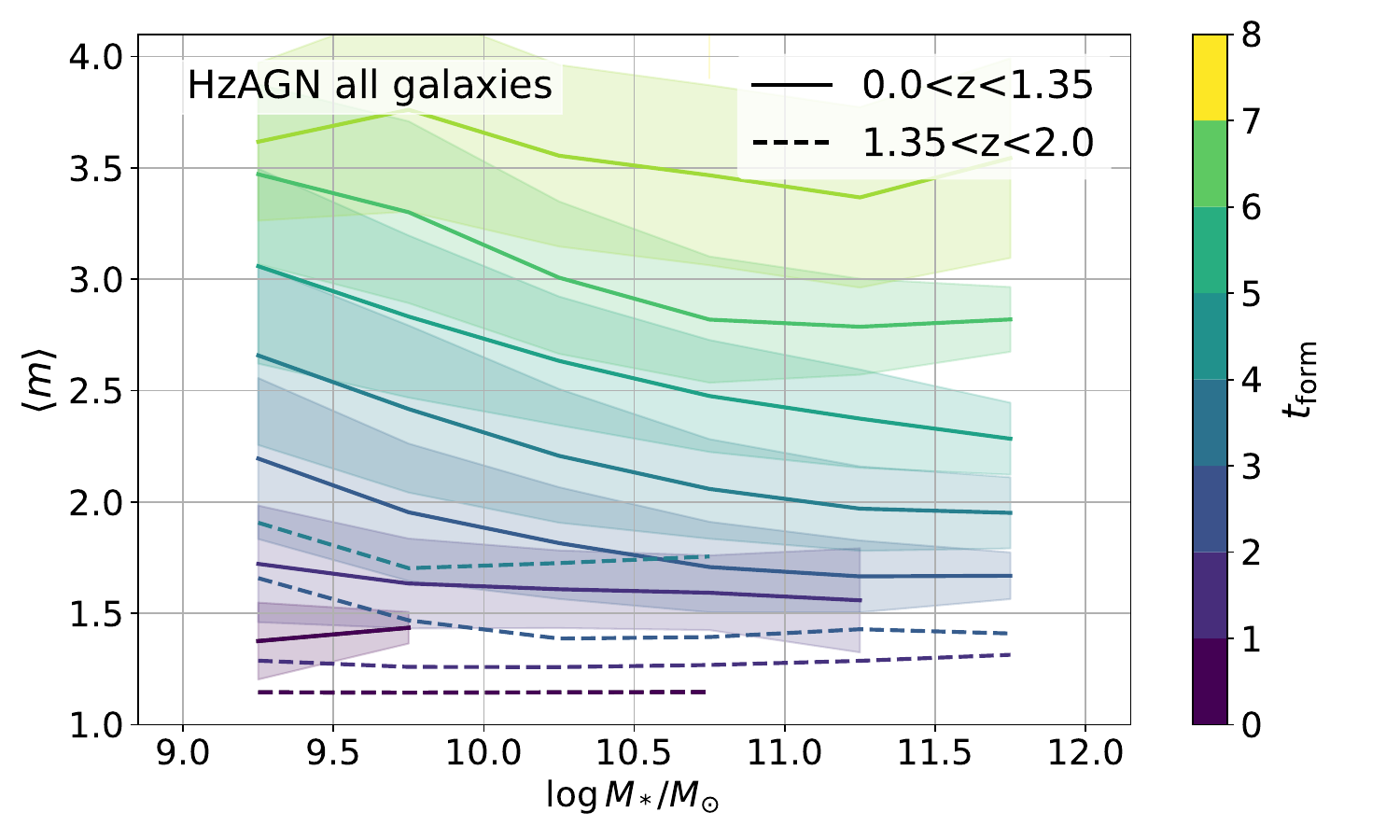}
\end{center}
\caption{Number of maxima  $\langle m \rangle$ in COSMOS ({top}) and Horizon-AGN ({bottom}) SFHs as a function of stellar mass, colour-coded by formation lookback time for two different redshift bins. The number of maxima is computed as described in Section~\ref{sec:nmaxima} and Fig~\ref{fig:method_summary}. }
\label{fig:nmax}
\end{figure}

Figure~\ref{fig:nmax} shows the number of maxima in the reconstructed normalised SFHs, as a function of the formation time $t_{\rm form}$ and the stellar mass, in COSMOS2020 (top) and {\sc Horizon-AGN} (bottom). The number of maxima is computed on 50 individual realisations from the posterior distribution and averaged per galaxy, as described in Section~\ref{sec:nmaxima} and Fig.~\ref{fig:method_summary}. 

We find that the number of maxima depends on both stellar mass and formation time. At a given mass, galaxies that formed the bulk of their mass a longer time ago are more likely to have experienced a second and possibly third peak of star formation, a trend found at all masses and both in COSMOS2020 and {\sc Horizon-AGN}. Furthermore, high-mass galaxies have a lower number of local maxima in their normalised SFHs than the least massive ones, suggesting that they assembled most of their mass at once, and the fraction of mass that could form in subsequent episodes is much lower (i.e., those late peaks do not meet our persistence criterion for peak identification). The trend is more pronounced in COSMOS2020 than in {\sc Horizon-AGN}. The phenomenon can also be interpreted as the accumulation of several disconnected star formation bursts in massive galaxies, potentially resulting in a general smoothing effect \citep[see also][for a similar intepretation based on the central limit theorem]{Iyer19}. 

With a different method to identify peaks, \cite{Iyer19} also found a mass dependency of the number of peaks in the SFH especially at low redshift ($z\sim 0.5$), with a higher probability for low-mass galaxies ($\log M_*/M_\odot < 10.5$) to exhibit several peaks compared to high-mass galaxies, which is in line with our findings. The trend they found is less clear at higher redshifts and possibly reverses, but we note that galaxies were not divided by $t_{\rm form}$ in their analysis. Finally, we emphasise that the mass dependency of the number of maxima in the SFH is particularly interesting considering that the mass was not an input of the reconstruction, since the algorithm is designed to infer SFHs from colours. 

Comparing the results in two different redshift bins, we notice that at fixed stellar mass and formation time, the average number of maxima is smaller at higher than lower redshift, especially for low-mass galaxies with small lookback formation time.  This suggests that the growth of galaxies at a lower redshift is less smooth than it is at higher redshift.

\begin{figure*}
\begin{center}
 \includegraphics[width=0.95\textwidth]{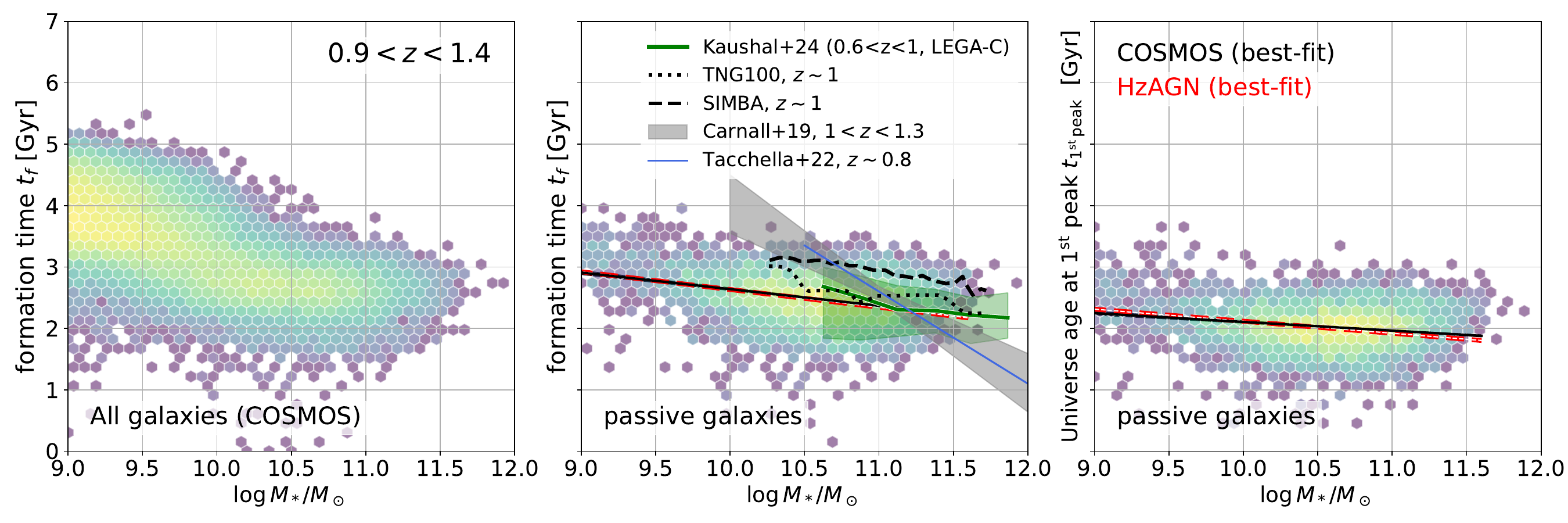}
 \caption{ Formation time $t_{\rm f}$ (i.e. the age of the Universe corresponding to the mass-weighted age of the galaxy as defined in equation~\ref{eq:tf}) versus mass in COSMOS2020 at $0.9<z<1.4$ compared to the literature for all galaxies (left) and passive galaxies only (middle). The right panel displays the age of the Universe corresponding to the first peak in the SFH. The black line is the best-fit for the passive galaxies in COSMOS2020, and the red line is the best-fit in {\sc Horizon-AGN}.}
  \label{fig:formationtime}
 \end{center}
\end{figure*}

\begin{figure*}
\begin{center}
 \includegraphics[width=0.99\textwidth]{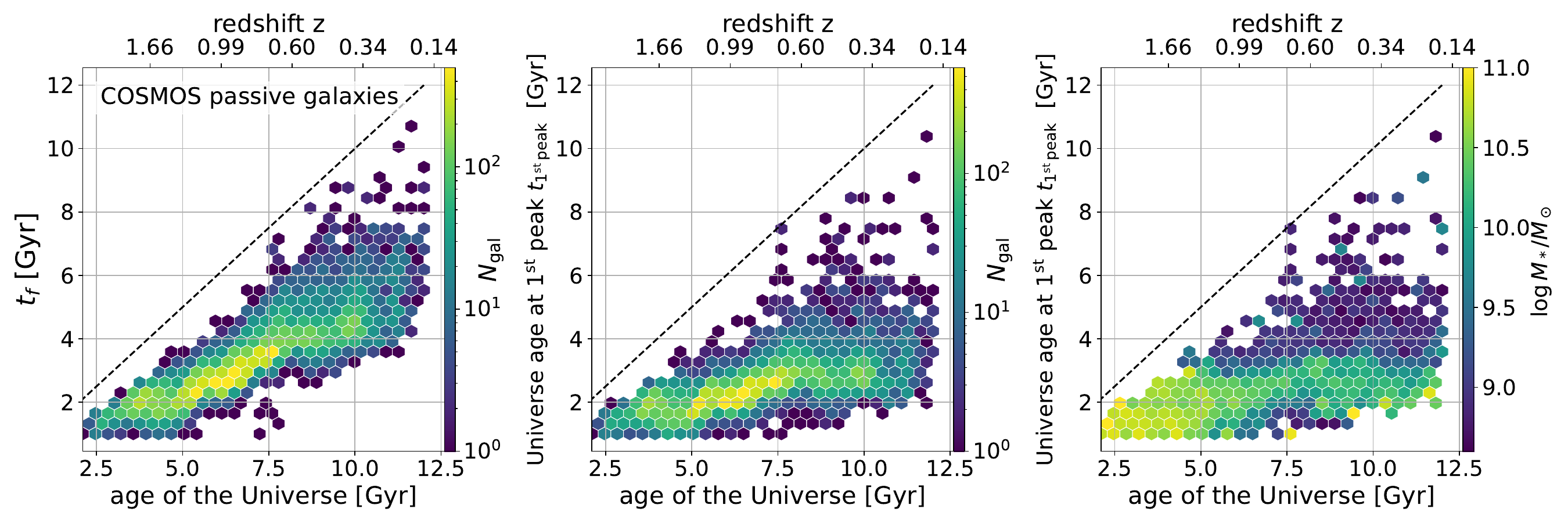}
 \caption{For passive galaxies in COSMOS2020, formation time $t_{\rm f}$  (left) and age of the Universe  at the first peak (middle and right) as a function of the redshift at which the galaxies are observed (or corresponding age of the Universe). Hexagons are colour-coded by density (left and middle) or stellar mass (right).}
  \label{fig:formationgal}
 \end{center}
\end{figure*}

\subsubsection{Formation redshift}
\label{sec:formationredshift}
Finally, we measure the formation time $t_{\rm f}$ (the age the Universe was at the mass-weighted age of the galaxy)  of all galaxies in the sample. To provide a direct comparison with existing observations, we first restrict our analysis to $0.9<z<1.4$, because data from the litterature are mostly confined to this redshift range. In particular, Fig.~\ref{fig:formationtime} shows the formation time  (the age of the Universe at which the galaxy age was equal to its mass-weighted age) as a function of stellar mass for all galaxies (left) and passive galaxies (middle), as well as the  the age of the Universe  at the first peak for passive galaxies (right). 
At a given stellar mass, galaxies exhibit a wide variety of formation times, with low-mass galaxies having formed the bulk of their mass after more massive ones.

We compare the formation time of passive galaxies with data from \cite{2024ApJ...961..118K} at $0.6<z<1$ in the LEGA-C sample (using their reconstruction based on the code {\tt Prospector} described in \citealp[]{{2021ApJS..254...22J}}, see also our comparison in Appendix~\ref{ap:legac}), data from TNG100 \citep{pillepich18} and SIMBA \citep{simba19} simulations\footnote{datapoints were extracted from \cite{2024ApJ...961..118K} with \href{https://plotdigitizer.com/app}{https://plotdigitizer.com/app} 
}, and the relation fitted by \cite{Carnall19} at $1<z<1.3$ and \cite{tacchella22} at $z\sim 0.8$. The estimated formation time of passive galaxies in both COSMOS2020 and {\sc Horizon-AGN} is very similar to \cite{2024ApJ...961..118K}, and spans the same range as \cite{{Carnall19}} and \cite{tacchella22}, but with little variation with mass. We note, however, that due to the sparsity of their sample, the slope of the relation fitted in these papers for galaxies $\log M_*/M_\odot >10.5$ might be very sensitive to the inclusion or not of the highest-mass galaxies ($\log M_*/M_\odot >11.5$), where the galaxy density is also the lowest. On the other hand, our outlier rejection algorithm tends to reject the most massive galaxies (see Fig.~\ref{fig:popoutliers2}), and therefore our sample could be incomplete at the high mass end, biasing the slope of this fit.

Following \cite{Gallazi14} at $z\sim 0.7$ and \cite{Carnall19} at $1<z<1.3$, we also fit the relation between the formation time $t_{\rm f}$ and the stellar mass. We perform this fit at $0.9<z<1.4$ to provide a direct comparison with existing observational data and we find, for the passive populations: 
\begin{equation}
\left( \frac{t_{\rm f}}{{\rm Gyr}}   \right) = 2.369 \pm 0.008-(0.271 \pm 0.010 )\times \log \left( \frac{M_*}{10^{11}M_\odot}\right) \, .
\end{equation}

We also fit the relation between the age of the Universe at the first peak $t_{\rm 1^{\rm st}\,peak}$ and the stellar mass, which provides an even flatter relation:
\begin{equation}
   \left( \frac{t_{\rm 1^{\rm st}\,peak}}{{\rm Gyr}}   \right) = 1.964 \pm 0.008-(0.142 \pm 0.010 )\times \log \left( \frac{M_*}{10^{11}M_\odot}\right)\,.
\end{equation}

In an effort to further understand the buildup of the population of passive galaxies, Fig.~\ref{fig:formationgal} shows the formation time (the age of the Universe at which the galaxy age was equal to its mass-weighted age) and age of the Universe at the first peak as a function of the age of the Universe $t_{\rm Universe}$ at which the galaxies were observed for all passive galaxies in COSMOS2020 (left and middle panels, respectively). We show only galaxies more massive than $\log M_*/M_\odot>8.6$, to guarantee completeness up to $z\sim 1.5$ \citep[see e.g.][]{shuntov22}. At an age of the Universe above 5\,Gyr ($z< 1.3$), the passive population lacks galaxies with formation times below $\sim 2$\,Gyr (formation redshift above $z_{\rm f}>3.3$), whereas galaxies with such formation time were found at higher redshift. This is possibly due to incompleteness in our sample due to the outlier rejection algorithm. It is also possible that these passive galaxies have rejuvenated and therefore moved to the star-forming population by $z\sim 1.3$, or that they experienced recent peak of mass assembly, which shifted their formation time towards more recent epochs.
In the middle panel, the position of the first peak only weakly depends on the redshift. However, the mean mass of the population at a given $\left(t_{\rm Universe},t_{1^{\rm st}\,{\rm peak}} \right)$ decreases, indicating the appearance of a new population of lower-mass galaxies. These galaxies would have formed almost as early as the first massive passive ones because the position of the first peak does not drastically change. However, they might have experienced star formation for a longer period of time, or multiple bursts, making them join the passive population later. 
Overall, these passive galaxies have experienced major episodes of mass assembly by $z\sim 3$ ( $1<t_{\rm Universe}/\rm Gyr<4$, i.e. $2.2<z<5.8$), whatever their mass. 
Finally, we note that from $t_{\rm Universe}\sim 7.5\,\rm Gyr$ ($z\sim 0.7$) the scatter of $t_{1^{\rm st}\,{\rm peak}}$ increases with the inclusion of very low-mass galaxies ($\log M_*/M_\odot < 9$) having recently formed, possibly satellite galaxies \citep[see also][about the build-up of the passive population in COSMOS2020]{weaver23}. These recently formed low-mass galaxies do not seem to represent an important fraction of the passive population at higher redshift ($z> 0.7$), but it is still possible that low-mass passive galaxies are missing from our sample at higher redshift due to the outlier rejection (see Fig.~\ref{fig:popoutliers}).

 \section{Caveats and perspectives}
\label{sec:caveats}
\subsection{Sensitivity to filter selection}
\label{subsec:dust_problem}

\begin{figure*}[h]
    \centering
    \begin{subfigure}
        \centering
        \includegraphics[width=0.48\linewidth]{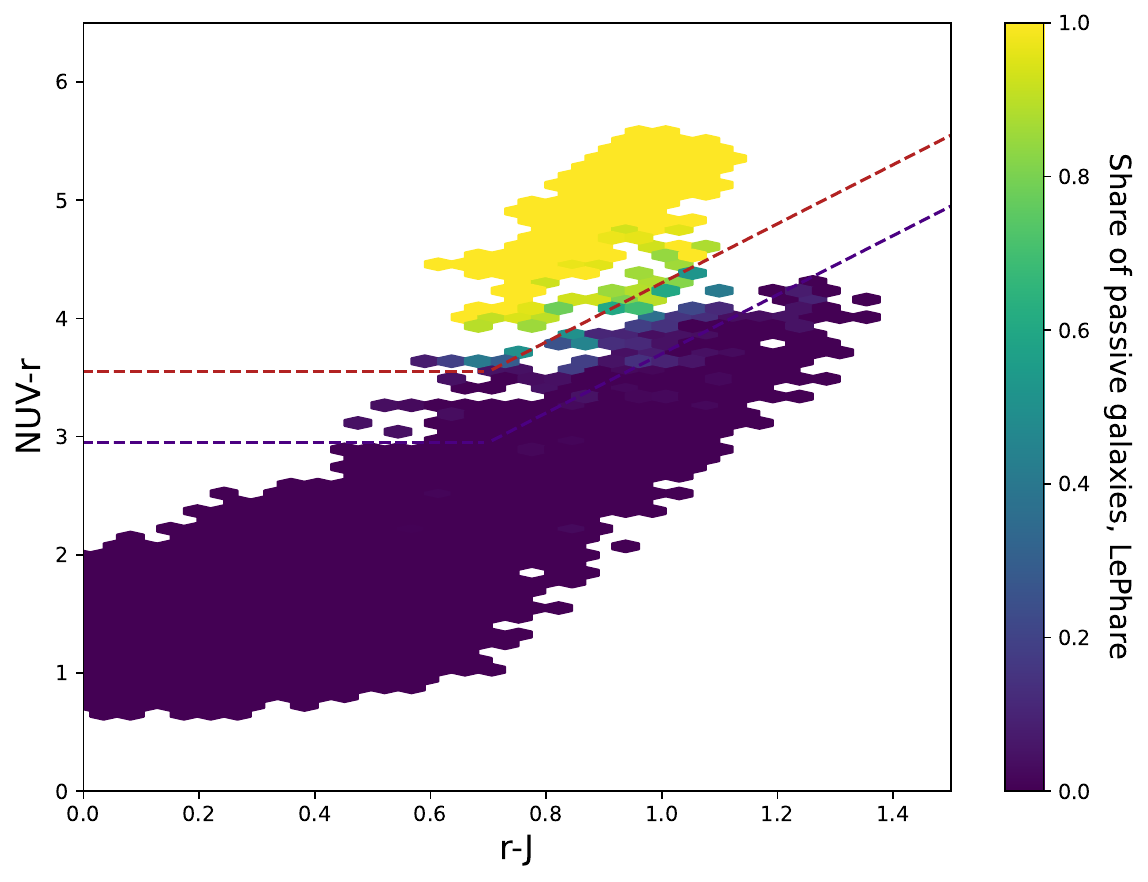}
    \end{subfigure}
    \hfill
    \begin{subfigure}
        \centering
        \includegraphics[width=0.48\linewidth]{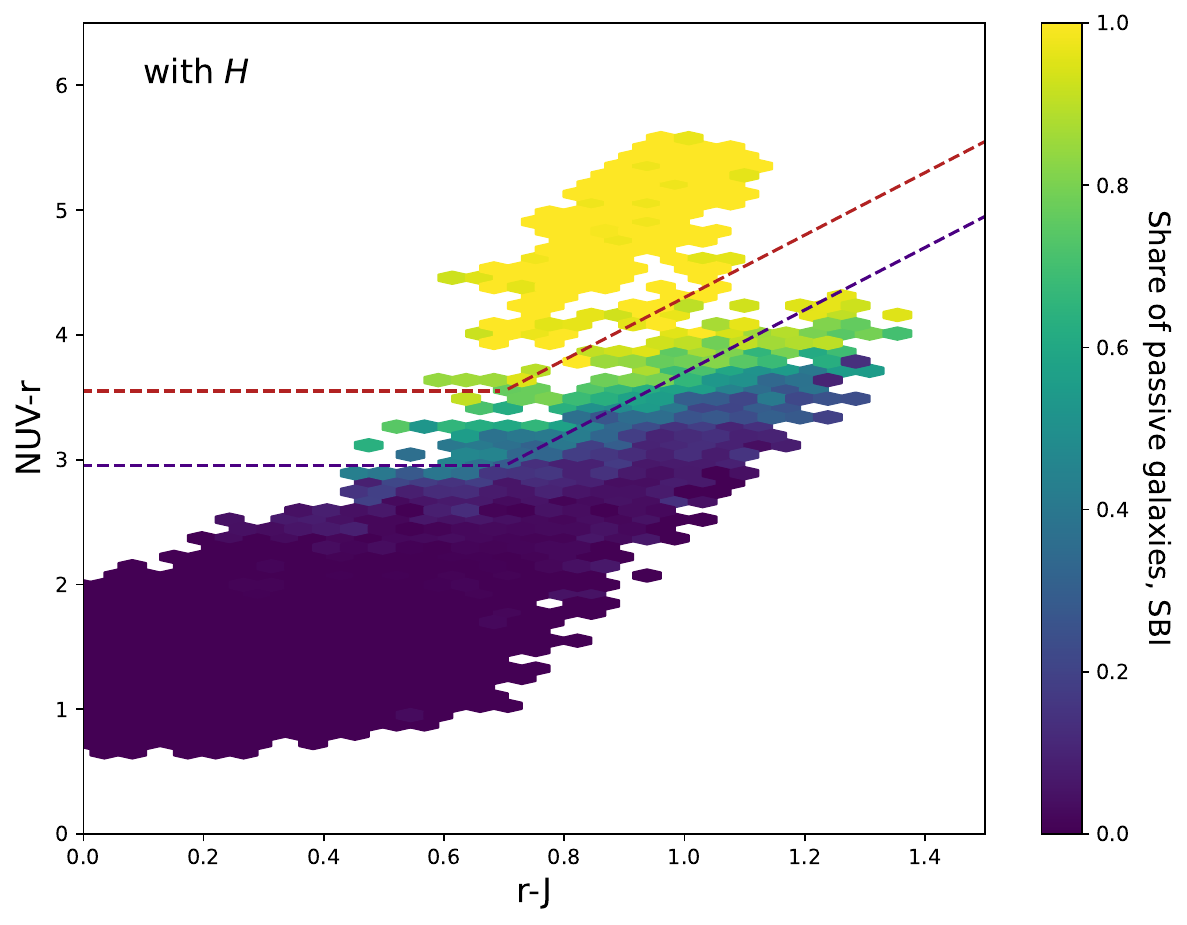}
    \end{subfigure}
    
    \vskip\baselineskip
    
    \begin{subfigure}
        \centering
        \includegraphics[width=0.48\linewidth]
        {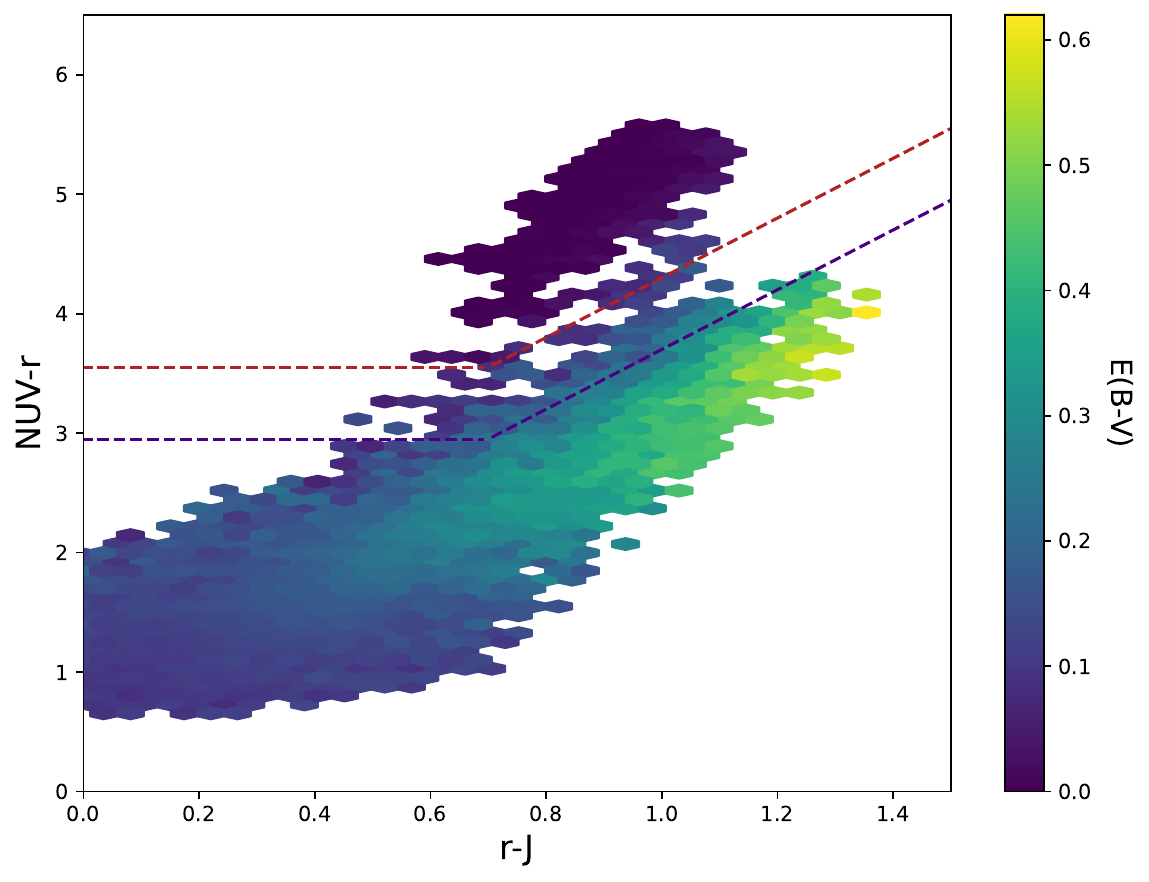}
    \end{subfigure}
    \hfill
    \begin{subfigure}
        \centering
        \includegraphics[width=0.48\linewidth]{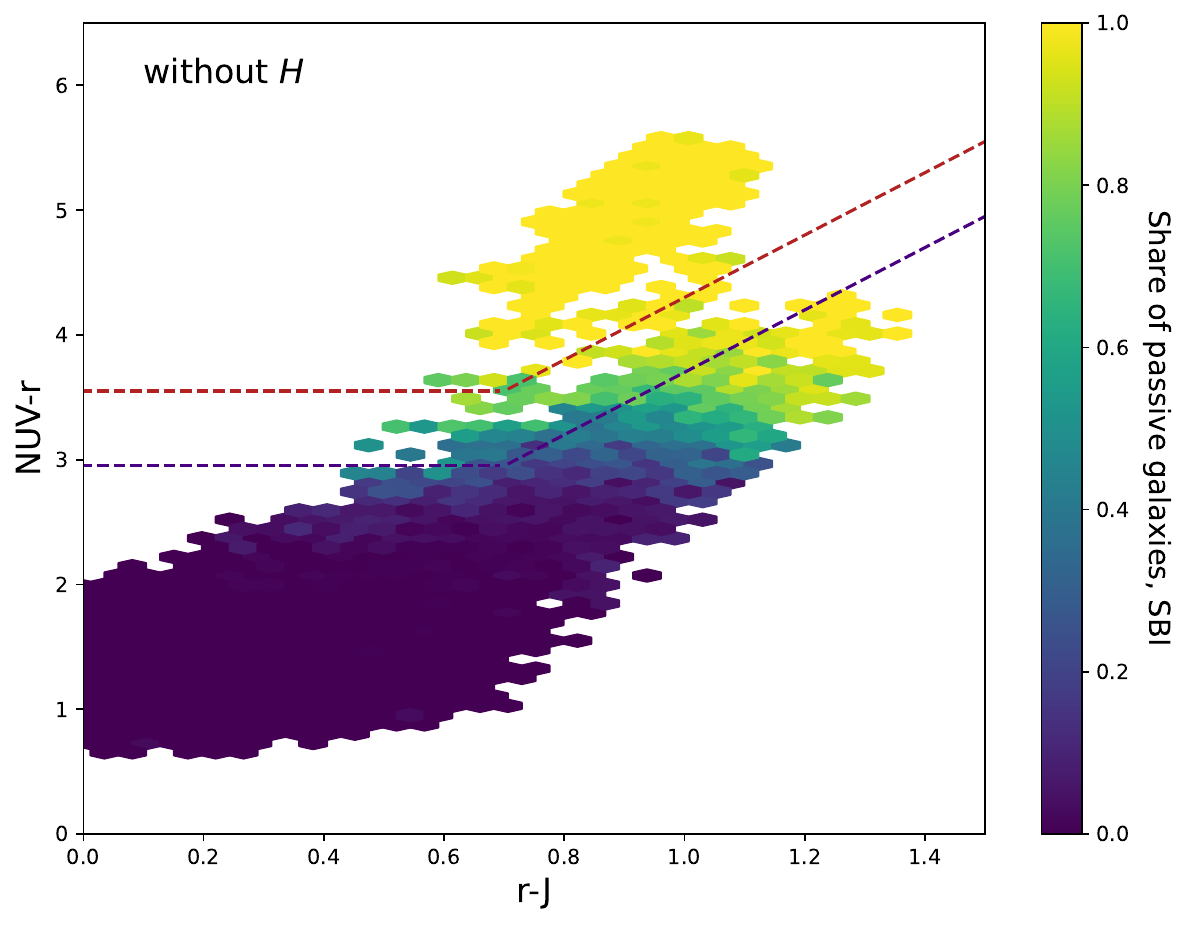}
    \end{subfigure}
    
    \caption{$\rm{NUV}\mathit{rJ}$ diagrams showing the separation of active and passive galaxies in $0.5<z<1$ in COSMOS2020. The dashed line is from \cite{ilbert13}. {Top left}: The fraction of passive galaxies (colours) selected as those with a \texttt{LePHARE} estimate of $\log(\mathrm{sSFR}/\rm yr^{-1})<-11$. {Top right}: The same fraction from sSFRs estimated by our SBI methodology using all filters. {Bottom left}: The distribution of dust, as estimated by \texttt{LePHARE}. {Bottom right}: The same fraction from sSFRs estimated by our SBI methodology but removing the $H$ Band.  The SFH inferred by SBI without the $H$ band on the right side of the panel corresponds to more passive galaxies than estimated with SED fitting, which identifies those galaxies as active but dusty.}
    \label{fig:NUVrJ_dust}
\end{figure*}

Section~\ref{sec:results} shows that our algorithm performs generally as expected (which a comparison with the literature in Appendix~\ref{ap:legac} confirms). However, as for all SED fitting, performances can vary and degeneracies appear depending on the chosen set of colors. For instance, we find a tendency for our network to inaccurately  classify some massive dusty star-forming galaxies as passive, a tendency which is amplified after removing the $H$-band. This can be seen in Fig.~\ref{fig:NUVrJ_dust} which shows the distribution of our $0.5<z<1$ sample in the $\rm{NUV}\mathit{rJ}$ diagram. 

The upper and middle panels illustrate an inconsistency between our SFHs and the \texttt{LePHARE} SFR estimates. According to \cite{ilbert13} and as illustrated in the top panel, passive galaxies with a specific SFR (sSFR) of $\log(\mathrm{sSFR}/\rm yr^{-1})<-11$ should be above the red line in the $\textrm{NUV}\mathit{rJ}$ diagram. However, when removing the $H$ band, our method finds some galaxies below the line with a passive SFH. Since this part of the diagram is populated by massive dusty galaxies (compared to the bottom panel), our estimator could be confusing red, dust-attenuated galaxies with passive galaxies.
We note that the perfect separation in the top panel is somewhat artificial: we use rest-frame magnitudes computed by \texttt{LePHARE}, and therefore we expect independent sSFR estimates to not match exactly this pattern.

\subsection{Is our {\sc Horizon-AGN} prior valid?}
\label{subsec:validity}
Our simulated SFHs are representative of real ones only if all internal (e.g., feedback, morphology) and external processes (interaction, mergers, matter acquisition from the cosmic web) shaping them are correctly accounted for in the simulation. Each effect is expected to act at a given epoch on a particular scale and at a particular timescale, and will imprint a specific signature in galaxy SFH and spectra. Our network is trained on the {\sc Horizon-AGN}  simulation, and therefore the SFHs that we can predict are those of the {\sc Horizon-AGN} galaxies. However, we know that {\sc Horizon-AGN} does not predict the correct abundances of the passive galaxy population (see Section~\ref{sec:hzagn}). Some SFH might therefore be missing, which could affect the correct prediction in observations. Nevertheless, this limitation is inherent to any SED-fitting code. We emphasise that our method enables higher-quality SFH recovery at a lower computational cost than classical SED fitting techniques.   
The analysis shown in Fig.~\ref{fig:slopecos} is reassuring. In fact, the trend of the SFH slope as a function of mass is different between COSMOS2020 and {\sc Horizon-AGN}, suggesting that the result of the reconstruction is not entirely determined by the prior.

Furthermore, it is important to note that by design we have rejected observed galaxies which have a photometry incompatible with the simulations (Section~\ref{sec:outlierdetectioncos}). The predictive power of our network is therefore limited to those galaxies which have a photometry similar the {\sc Horizon-AGN} simulation, and the analysis  of  the sample of passive galaxies might be partially incomplete (see e.g. the discussion about Fig.~\ref{fig:formationgal} in Section~\ref{sec:formationredshift}).
A possible improvement of our method would be to enrich the training set with photometry derived from other hydrodynamical cosmological simulations or semi-analytical models. 

\subsection{Disentangling between in-situ and ex-situ mass assembly}
One particularity of our method is that we reconstruct the history of mass assembly instead of the history of star formation. Therefore we cannot disentangle between in-situ and ex-situ stellar population \citep[see e.g.][for a discussion]{2024ApJ...961..118K}. For those galaxies that underwent mergers, we reconstruct in fact the cumulative SFH of the different progenitors that are part of the final passive galaxies. One possible improvement would be to try learning  simultaneously the merger and star-formation histories of the galaxies with our SBI method from their photometry.

\section{Conclusion}
\label{sec:conclusion}
In this paper, we presented a novel method to perform Bayesian estimation of the star formation history of galaxies from photometry using simulation-based inference with {\sc Horizon-AGN} hydrodynamical simulation together with a suitable parameterisation of the SFH. We showed that the method was able to correctly reconstruct the entire SFH of galaxies on simulations. We applied it to the COSMOS2020 catalogue. Our results are as follows:
\begin{itemize}
   \item Using simulated data, SBI is able to properly estimate the entire SFH, and accurately quantify the Bayesian uncertainties (Fig.~\ref{fig:perf_SBI});
    \item The shapes of our SFHs broadly agree with $\mathrm{NUV}\mathit{rJ}$ diagram classification of passive galaxies (Figs.~\ref{fig:activepassive} and~\ref{fig:NUVrJ_patches}). In addition, the estimates of the time of first stellar formation of galaxies are not in disagreement with age of the Universe at the corresponding redshift despite not using any information but the colours as input.  Although mass is not used as input, median SFHs of star forming galaxies exhibit on average a wide diversity depending on mass and redshift  while the median SFH of passive galaxies have very limited range of histories, with almost no dependence on the mass; 
    \item We have extracted summary statistics (slope, number of maxima and formation time) from the SFH of individual galaxies (Fig.~\ref{fig:method_summary}). The slopes of the SFHs of passive galaxies show only a weak trend with stellar mass at $z<1.35$ but a significant scatter, indicating that other parameters than mass could drive quenching. On the other hand, star-forming galaxies show a clear mass dependence, with lower-mass galaxies undergoing more vigorous recent star formation. Overall, galaxy slopes in COSMOS2020 vary over a wider range than in {\sc Horizon-AGN} (Fig.~\ref{fig:slopecos}).
    \item  Low-mass galaxies exhibit a larger number of peaks in their history of mass assembly than high-mass ones, and the trend is more evident in COSMOS2020 than in {\sc Horizon-AGN} (Fig.~\ref{fig:nmax}).
    \item At a given mass, we find a large diversity of formation redshift, but for passive galaxies the dependency of the formation redshift on mass is weak (Fig.~\ref{fig:formationtime}).  Most passive galaxies with $\log M_*/M_\odot > 9$ had a first event of mass assembly around $z\sim 3$ ($2.2<z<5.8$), disregarding their mass. From $z\sim 0.7$ some $\log M_*/M_\odot < 9$ which formed more recently joins the passive population (Fig.~\ref{fig:formationgal}).
\end{itemize}

\section{Acknowledgements}
CL thanks the Programme National Galaxies and Cosmologie (PNCG) for funding. GA thanks the Initiative Physique des Infinis (IPI) for supporting his research at IAP. The authors acknowledge the support of the French Agence Nationale de la Recherche (ANR), under grant ANR-22-CE31-0007 (project IMAGE). This project has received financial support from CNRS and CNES through the MITI interdisciplinary programs. This work was made possible by utilising the CANDIDE cluster at the Institut d’Astrophysique de Paris. The cluster was funded through grants from the PNCG, CNES, DIM-ACAV, the Euclid Consortium, and the Danish National Research Foundation Cosmic Dawn Center (DNRF140). It is maintained by Stephane Rouberol. LC acknowledges support from the French government under the France 2030 investment plan, as part of the Initiative d’Excellence d’Aix-Marseille Université – A*MIDEX AMX-22-RE-AB-101.
\bibliographystyle{aa}
\bibliography{aanda,hjmccrefs}

\appendix
\section{General validation of the reconstructed SFHs against the literature}
Here we investigate to what extent our recovered SFHs agree with previously published SFHs in COSMOS.

\subsection{The LEGA-C galaxy sample}
\label{ap:legac}

\begin{figure}
 \includegraphics[width=\columnwidth]{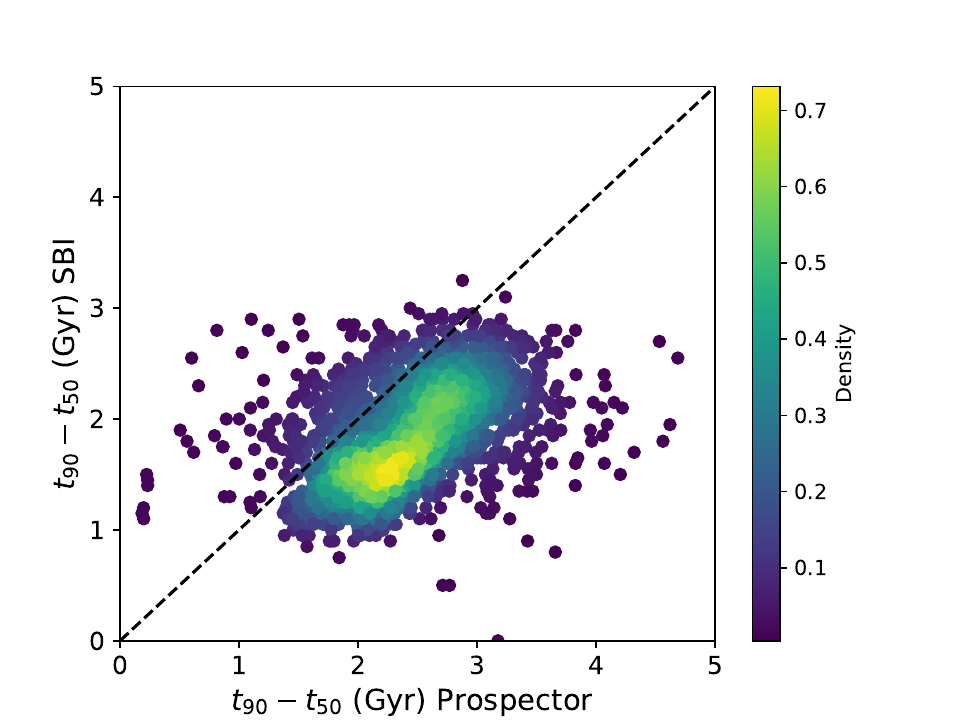}
  \caption{Comparison of $t_{90}-t_{50}$ estimated by our SBI method and {\tt Prospector} in  \citep{2024ApJ...961..118K}, for all galaxies in the selected sample described in Appendix~\ref{ap:legac}. }
    \label{fig:Prospector-sbi-bagpipes}
\end{figure}
\begin{figure}
 \includegraphics[width=\columnwidth]{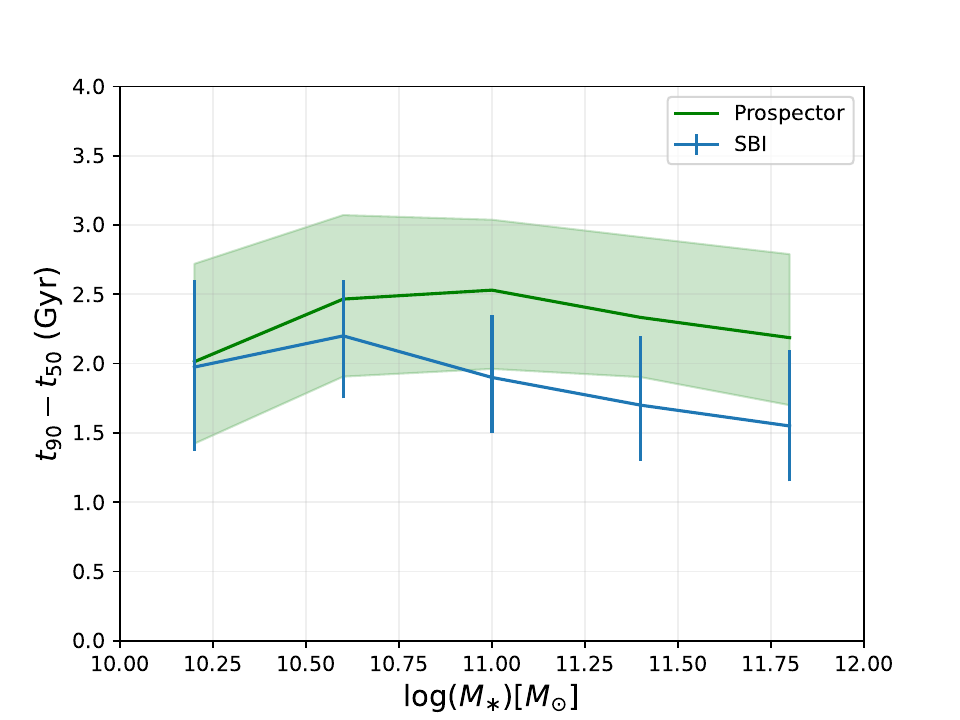}
  \caption{Variation of $t_{90}-t_{50}$ as a function of mass, estimated with \texttt{Prospector} (green) by \cite{2024ApJ...961..118K}, and with our SBI method (blue). Error bars and shaded regions represent the 16th-84th confidence intervals.}
  \label{fig:mass_kaushal}
\end{figure}

We compare our SFHs with \cite{2024ApJ...961..118K} who use SED fitting using LEGA-C \citep{2021ApJS..256...44V} spectroscopy as well as photometry to perform SFH inference for $\sim3000$ massive galaxies at $0.6<z<1$. They compare fits obtained by {\tt Bagpipes} and {\tt Prospector} codes \citep{2021ApJS..254...22J} which use respectively a double power law and a non-parametric SFH model. {\tt Prospector}'s SFH is using a continuity prior \citep{2020ApJ...904...33L} where the SFHs piecewise constant  with a Student’s t-distribution modelling the change in $\log {\rm SFR}(t)$ in adjacent time bins. Since our goal is to compare the reconstructions and not to confirm their scientific findings, we use the same galaxy sample without regard for completeness in mass or weighting corresponding to volume correction \citep{2024ApJ...961..118K, van_der_Wel_2021}. We downloaded the dataset provided by the authors\footnote{\url{https://simbad.u-strasbg.fr/simbad/sim-ref?querymethod=bib&simbo=on&submit=submit\%20bibcode&bibcode=2024ApJ...961..118K}}. In order to have a single data set for all comparisons, we selected the subset of galaxies for which: (i) {\tt Bagpipes} and {\tt Prospector} reconstructions are available; (ii) Our assumption $SNR>2$ is verified in the $K_{\rm s}$ band; (iii) Our outlier detection algorithm does not reject the galaxy .

After these cuts, $2200$ individual galaxies are left. For each, we reconstruct $t_{10}, t_{50}$ and $t_{90}$, corresponding to the times at which the galaxy has formed $10, 50$ and $90$ percent of its mass respectively. We follow the authors in focussing on the $t_{90} -t_{50}$ statistics. The top panel of Fig.~\ref{fig:Prospector-sbi-bagpipes} shows the $t_{90}-t_{50}$ estimates measured using our SBI reconstruction compared to the same estimate from {\tt Prospector}, for all galaxies in the sample. It shows that our SBI measurements are in broad agreement with {\tt Prospector}'s, despite systematic underestimates. However, this discrepancy remains relatively small given the overwhelming differences in methodology and modelling assumptions.

Finally, in Fig.~\ref{fig:mass_kaushal} we show the mass dependency of the $t_{90} - t_{50}$ statistics, as also presented in   \cite{2024ApJ...961..118K}, and we confirm that both methods are compatible with a constant $t_{90} - t_{50}$  mass range $10.25 < \log M_*/M_\odot < 12$  for this small sample of galaxies.

\subsection{The \cite{2024arXiv240417945P} galaxy sample}

\begin{figure*}
 \includegraphics[width=\textwidth]{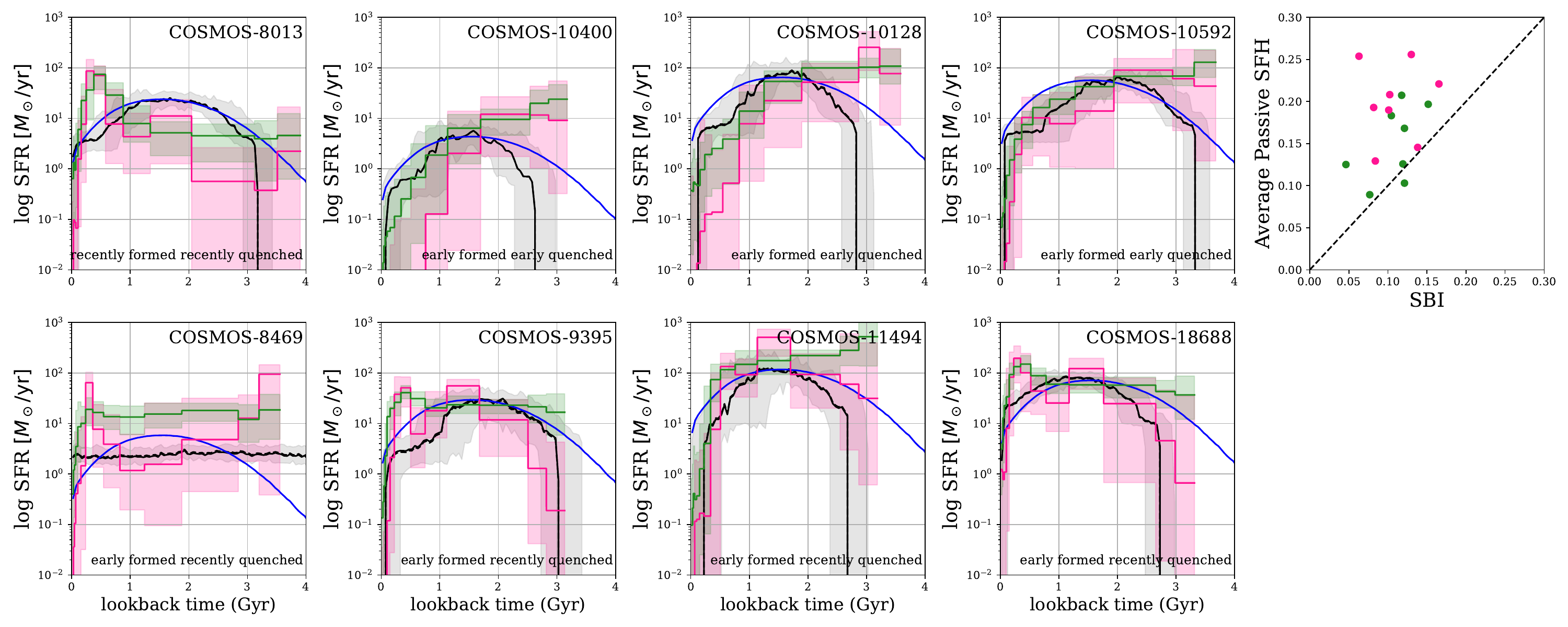}
 \caption{Comparison of SFH estimates for individual galaxies in the \cite{2024arXiv240417945P} sample. The shaded green regions shows the results using the continuity prior, in pink using the bursty prior, and in black our SBI estimate. The blue line shows the average SFH of passive galaxies. The size of the shaded regions corresponds to the $68\%$ confidence intervals. The top right panel shows the median residuals between the the \cite{2024arXiv240417945P} estimates and either our SBI estimate (horizontal axes) or the average passive SFH (vertical axes). The points above the diagonal line indicate our method is closer to \texttt{Prospector}'s estimate than the average passive SFH is. }
 \label{fig:park}
\end{figure*}

We now compare our estimates with individual galaxies in \cite{2024arXiv240417945P}. They studied 151 galaxies in the COSMOS field using data from the Blue Jay survey, a JWST Cycle 1 program (GO 1810; PI: Belli). After performing a spectro-photometric SED fit using {\tt Prospector} on each galaxy, they selected 16 massive quiescent galaxies for which they have both NIRSpec spectroscopy (and spectroscopic redshifts) and photometry from HST/ACS+WFC3 and IRAC bands from \textit{Spitzer}. Massive galaxies were selected with a cut at $ \log M*/M_\odot > 10.0$ and as quiescent with an sSFR one dex below the star forming main sequence from \cite{leja22}. They carried out SED fitting with two different SFH priors : (1) the `continuity'  prior which uses a Student-T distribution with $\sigma = 0.3$ and $\nu = 2.0$  \citep{leja19a}, and (2) the `bursty' prior from \citep{tacchella22}, with $\sigma = 0.1$ which allows larger changes in star formation between two consecutive time bins.

We identified the same 16 galaxies in COSMOS2020. We rejected eight of them for either missing one of our photometric bands or being rejected by our outlier detection pipeline. Because our aim is to assess the performance of our method using a specific set of assumptions, we simply remove these galaxies and compare the remaining eight galaxies\footnote{We note that if studying those specific galaxies was our goal, we would only need to tune the training set of our network to account for the missing bands.} 

The green lines in Fig.~\ref{fig:park} shows the SFH estimated by \citep{2024arXiv240417945P} using the continuity prior in green and the bursty prior (in pink) compared  to our reconstruction ({black}). Since our method estimates only the \textit{normalized} SFH, we used the stellar mass from \citep{weaver22} to rescale them. We observe that the SFHs recovered with our method overall share the same shape across the sample (except for the peculiar COSMOS-8469), which is consistent with the selection of quenched galaxies and the homogeneity of massive quenched galaxies' SFH described in \ref{subs:mass_type}. 

Our estimates agree broadly with \citep{2024arXiv240417945P} despite our smaller credible intervals. In particular, COSMOS-8469's recovered SFH is always consistent with a constant SFH, despite the overall stellar mass appearing to vary significantly between fits. The galaxy formation ages are broadly consistent despite the fact that no redshift information is used in our fit. Interestingly, our method correctly estimates the age of formation of COSMOS-10400 at around 3\,Gyr despite a significant difference between the spectroscopic redshift ($z\sim 2.1$) and the photometric redshift estimated by \texttt{LePHARE} ($z\sim 0.99$).

Finally, we compute the average SFHs of passive galaxies in the range $1.6<z<2.7$ (in blue). We compare this naive estimate from our SBI one with {\tt Prospector} by computing their respective distances; this is shown on the top right panel of Fig. ~\ref{fig:park}. In each case, for SBI, we computed the median distance between {\tt Prospector}'s estimates and 50 draws in the posterior distribution. This shows that SBI is consistently closer to {\tt Prospector} than the naive estimate, which is particularly striking for the peculiar COSMOS-8469 object.

\section{Quality assessment of the summary quantities build from the SFH}
\label{ap:qa}
\begin{figure*}
\begin{center}
 \includegraphics[width=0.95\textwidth]{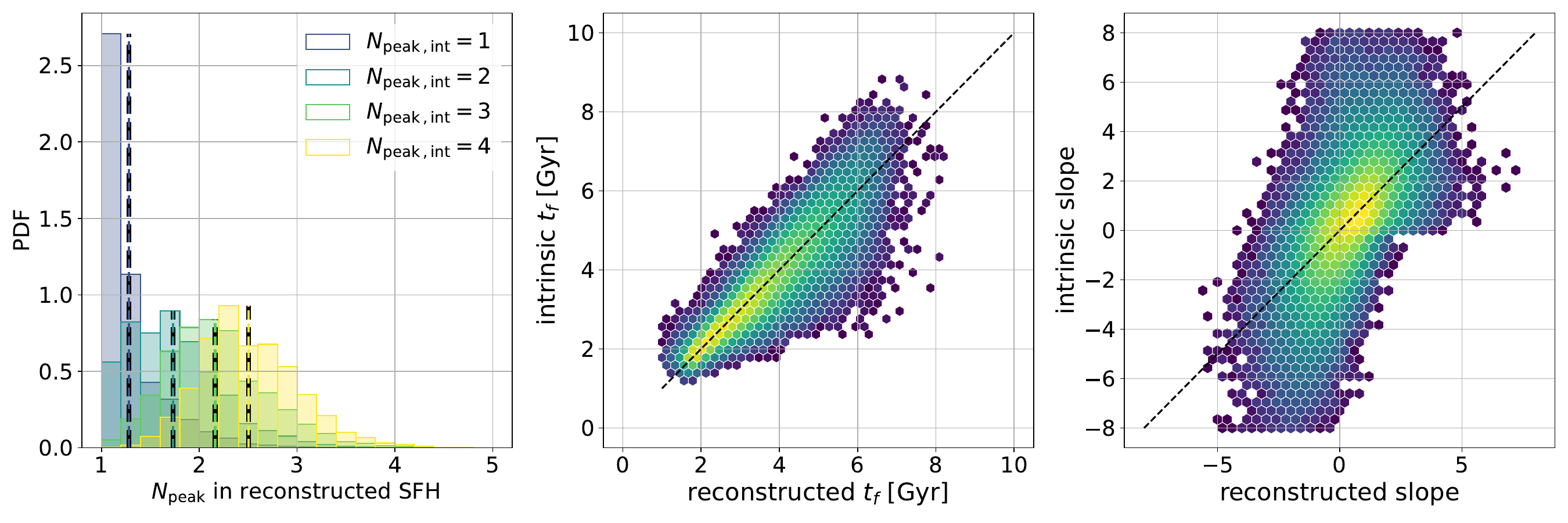}
 \caption{Comparison of the number of peaks, formation time and slope on reconstructed SFHs and intrinsic SFHs in the {\sc Horizon-AGN} simulation. For the reconstructed SFHs, the summary statistics are measured on 50 individual realisations from the posterior distribution and then averaged.}
  \label{fig:qasfh}
 \end{center}
\end{figure*}

In Section~\ref{sec:slope} we presented several summary statistics from the SFH: the slope, the number of peaks and the formation time (see Fig.~\ref{fig:method_summary}). 
To assess the quality of these estimators, Fig.~\ref{fig:qasfh} shows the comparison between these quantities derived from the reconstructed SFHs in {\sc Horizon-AGN} and those measured on the `true' SFHs in the simulations. For the reconstructed SFH, the summary statistics are measured on 50 individual realisations from the posterior distribution, and then averaged. The mean number of peaks is generally biased towards lower values, although the hierarchy is correctly recovered. The reconstructed and intrinsic formation time and slope are reasonably well correlated.

\begin{figure*}
\begin{center}
  \includegraphics[width=0.9\textwidth]{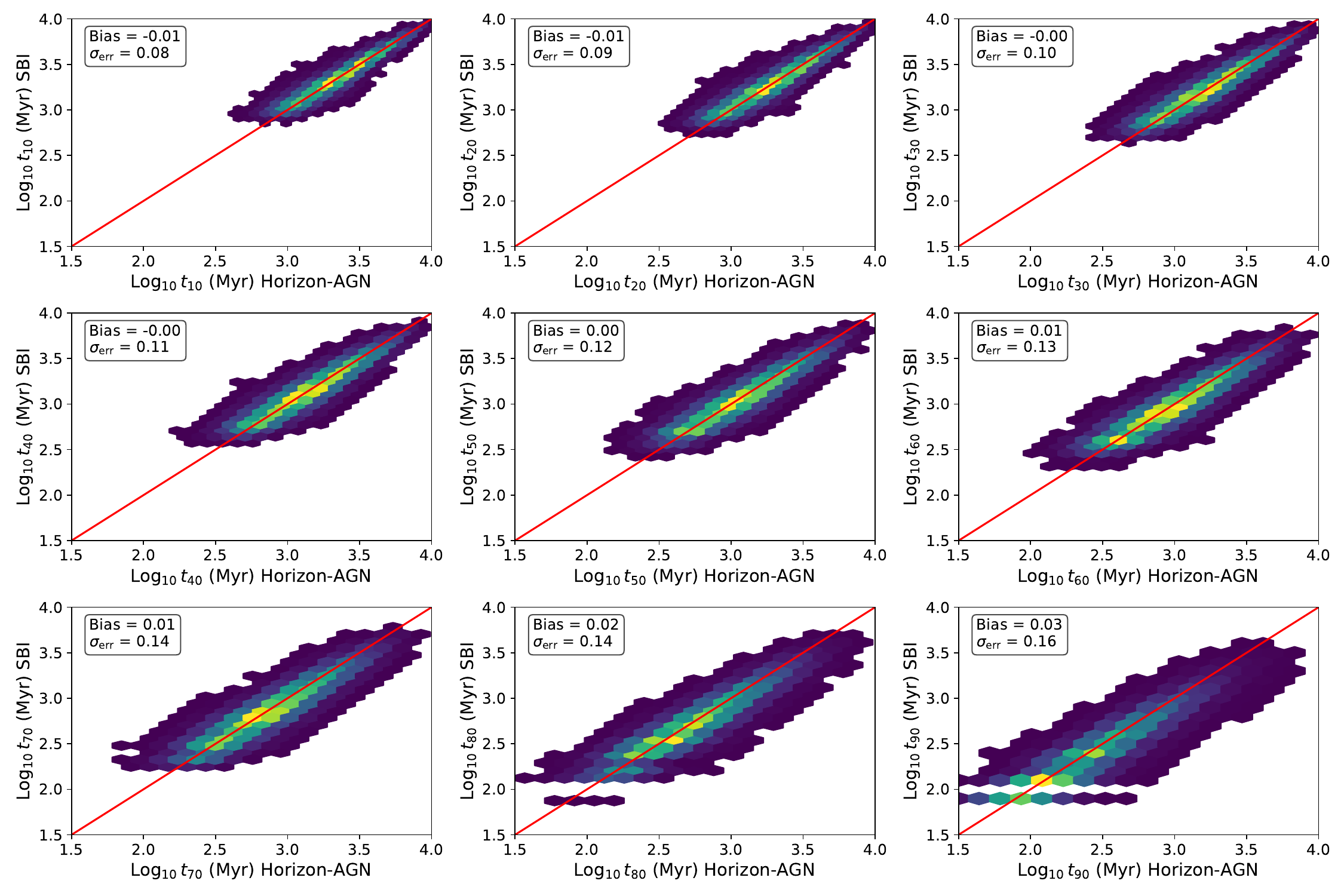}
  \caption{Comparison between the true formation times and our estimates, decile by decile, on $\sim 85000$ galaxies randomly drawn in the Horizon-AGN simulation and not used in the training of the neural network. }
  \label{fig:quantiles}
\end{center}
\end{figure*}
\label{lastpage}
\end{document}